\newif\ifPDF
\ifx\pdfoutput\undefined\PDFfalse
\else\ifnum\pdfoutput > 0\PDFtrue
     \else\PDFfalse
     \fi
\fi
\documentclass[usenatbib]{mn2e}
\ifPDF
  \usepackage[T1]{fontenc}
  \usepackage{aeguill}
  \usepackage[pdftex]{graphicx,color}
  \usepackage{subfigure}
  \usepackage{afterpage}
\else
  \usepackage[T1]{fontenc}
  \usepackage[dvips]{graphicx,color}
  \usepackage{subfigure}
  \usepackage{afterpage}
\fi

\newcommand{\Teff}{\ensuremath{T_{\rm eff}}}
\newcommand{\logg}{\ensuremath{\log{g}}}

\newcommand\5{{\footnotesize V}}
\newcommand\4{{\footnotesize IV}}
\newcommand\3{{\footnotesize III}}
\newcommand\2{{\footnotesize II}}
\newcommand\1{{\footnotesize I}}
\newcommand\lam{{$\lambda$}}

\newcommand{\ph}[1]{\phantom{#1}}

\newcommand{\ie}{{i.e.}}

\newcommand{\kms}{{km$\;$s$^{-1}$}}

\newcommand{\pA}{\ensuremath{\phantom{1}}}
\newcommand{\ppB}{\ensuremath{\phantom{+16}}}
\newcommand{\sref}[1]{Section~\ref{#1}}

\newcommand{\BJ} {\ensuremath{B_{\rm J}}}
\newcommand{\Ha} {\ensuremath{\rm{H}\alpha}}
\newcommand{\Hb} {\ensuremath{\rm{H}\beta}}
\newcommand{\Hg} {\ensuremath{\rm{H}\gamma}}
\newcommand{\Wl} {\ensuremath{\rm{W}_\lambda}}
\newcommand{\HeI}{\rm{He$\;$\sc{i}}}

\newcommand\mg{\ensuremath{^{\rm m}}}

\title[A 2dF survey of the SMC]
{A 2dF survey of the Small Magellanic Cloud}

\author[C. J.~Evans, I. D.~Howarth et al.]
{Christopher J.~Evans$^1$\thanks{email: cje@ing.iac.es;
idh@star.ucl.ac.uk;
mike@ast.cam.ac.uk;
awxb@star.ucl.ac.uk;
th@astro.ex.ac.uk},
Ian D.~Howarth$^2$,\newauthor
Michael~J.~Irwin$^3$, 
Adam~W.~Burnley$^2$ 
\& Timothy~J.~Harries$^4$
\\
$^1$Isaac Newton Group of Telescopes, Apartado de Correos 321, 
38700 Santa Cruz de la Palma, Canary Islands, Spain \\
$^2$Department of Physics and Astronomy, University
College London, Gower Street, London WC1E~6BT,~UK \\
$^3$Institute of Astronomy, University of Cambridge, Madingley Road, 
Cambridge CB3 0HA,~UK\\
$4$School of Physics, University of Exeter, Stocker Road, Exeter EX4 4QL,~UK}

\date{Received:}

\begin{document}
\maketitle

\begin{abstract}
We present a catalogue of new spectral types for hot, luminous stars
in the Small Magellanic Cloud.  The catalogue contains 4161 objects,
giving an order of magnitude increase in the number of SMC stars with
published spectroscopic classifications.  The targets are primarily B-
and A-type stars (2862 and 853 objects respectively), with 1
Wolf-Rayet, 139 O-type, and 306 FG stars, sampling the main sequence
to $\sim$mid-B.  The selection and classification criteria are
described, and objects of particular interest are discussed, including
UV-selected targets from the Ultraviolet Imaging Telescope (UIT) experiment, Be
and B[e] stars, `anomalous A super\-giants', and composite-spectrum
systems.  We examine the incidence of Balmer-line emission, and the
relationship between \Hg~equivalent width and absolute magnitude for BA stars.

\end{abstract}


\section{Introduction}
\label{intro}
Since the seminal study of the solar-neighbourhood stellar initial
mass function (IMF) by \citet{s55}, there have been numerous
observational and theoretical studies of the IMF in the Galaxy, the
Magellanic Clouds, and beyond
\citep[{cf.}][]{imfconf}.  One of the principal tools used in the
investigation of the IMF is the comparison of observed
Hertzsprung-Russell diagrams (HRDs) with those obtained from
population syntheses built on stellar-evolution models.  The
construction of observational HRDs for large samples of roughly
equidistant stars is most easily accomplished in the colour--magnitude
plane.  However, optical observations only sample the Rayleigh-Jeans
tail of the spectral energy distribution of O- and early B-type stars;
thus for these objects optical photometric colours alone do not
provide adequate discrimination in temperature (nor, therefore, in
luminosity) for satisfactory transformation between the observed and
theoretical HRDs.

The Small Magellanic Cloud (SMC) is of particular interest as the
nearest system with a well-established, substantial (factor $\sim$5)
underabundance of metallic elements compared to those found in the
Milky Way.  This makes it an attractive target for investigating the
role of metallicity in star formation and evolution, as well as a
variety of other topics in stellar and galactic astrophysics.  With
this in mind, we have undertaken a spectroscopic survey of the SMC's
hot, luminous stars, using the multi-fibre 2-degree Field (2dF)
instrument of the Anglo-Australian Telescope (AAT), primarily in order to
investigate directly the massive-star IMF.

The basic data of the 2dF spectroscopic survey are presented in this
paper, wherein catalogue entries are identified by `2dFS' numbers for
convenience.  Target selection, observations and data reduction are
described in
\sref{data}, while the criteria used for spectral and luminosity
classification of the sample are discussed in Sections~\ref{class}
and~\ref{chx_sx}.  Some aspects of the photo{\-}metry are discussed in
Sections~\ref{sec_phot} and~\ref{MBWL}.  Observations of stars of
particular interest are presented in \sref{seren}, and
catalogue contents are
summarized in Appendix~\ref{app_cat}.  An investigation
of the IMF of the SMC, based in part on the catalogue, will be presented
elsewhere.

\section{Data acquisition}
\label{data}

\subsection{Target selection: input catalogue}

\label{selection}

Our initial aim was to use photo{\-}metry to isolate those targets for
which spectroscopy was required for accurate placement in the
HRD.  The actual targets were then to be drawn from this
input catalogue.

At the start of the project, the best available source for both
large-scale photo{\-}metry and accurate astrometry (essential for the 2dF
observations) was scans of UK~Schmidt {\BJ}\footnote{\BJ\ 
indicates the use of Eastman Kodak IIIa-J emulsion ($+GG495$ filter);
$\BJ \simeq B - 0.28 (B - V)$ \citep{bg82}.}
and $R$ photographic survey plates made with the Automatic Plate
Measuring (APM) machine. Field \#29 in the Schmidt survey covers the greater
part of the SMC and was used to compile the input catalogue of
potential targets.  The densest regions (e.g., NGCs 330 and 346) were
too crowded for reliable APM scans and are largely excluded from our work.

For our investigation of the upper part of the IMF (and to avoid
contamination by foreground stars) the APM targets making up the input
catalogue for spectroscopy were restricted to bright blue stars, with
cuts of $({\BJ}-R)~<~0.1$ (corresponding, notionally, to O and B
spectral types) and ${\BJ} < 17.5$.  As the survey and analysis
progressed, it became evident both that additional selection effects
were in play (most importantly, systematic rejection of the brightest
stars as `non-stellar'), and that the photographic photo{\-}metry had
rather larger uncertainties than anticipated.  These additional
sampling effects will be discussed in our forthcoming analysis of the
IMF (and a more detailed
discussion of photo{\-}metry is given in Section~\ref{sec_phot}),
but one consequence is that the colour cut-offs were not as
effective as expected in isolating only the hottest stars.
As a result, some
relatively red stars were included in the observed 2dF sample,
including foreground objects.
(The expected surface density of foreground blue stars
is negligibly small -- about 1 per square degree.)
Fortunately, although our 2dF observations were not generally
optimized for radial-velocity measurements, they are quite sufficient to
discriminate between typical Galactic and SMC velocities
($\sim$+170~\kms).  Radial-velocity measurements identified 171 stars, 
mostly in the spectral range G0--K3,
as probable foreground objects.
These targets were discarded for the purposes of the spectroscopic
catalogue.

\begin{table}
\begin{center}
\caption{2dF field centres.  The fields were
observed in 1998 (September 25--28, fields 1--12) and 1999
(September~30/October~1, fields 13--18).}
\label{fields}
\begin{tabular}{c|c|c|c}
\hline
Field& \multicolumn{2}{c|}{Field centre (J2000)} & Integration \\
no. & $\alpha$ & $\delta$ & time (min)  \\
\hline
$\ph{1}$1 & 01 21 16.7 & $-$73 14 19 & $\ph{1}$60 \\
$\ph{1}$2 & 01 21 16.7 & $-$73 14 19 & 120  \\
$\ph{1}$3 & 00 41 53.2 & $-$73 33 33 & 150  \\
$\ph{1}$4 & 00 41 53.2 & $-$73 33 33 & 150  \\
$\ph{1}$5 & 01 06 32.4 & $-$72 43 58 & $\ph{1}$90 \\
$\ph{1}$6 & 01 06 31.1 & $-$73 03 58 & 120  \\
$\ph{1}$7 & 00 48 47.9 & $-$73 03 39 & 120  \\
$\ph{1}$8 & 01 06 31.1 & $-$73 03 58 & 150  \\
$\ph{1}$9 & 01 26 12.1 & $-$73 14 27 & 120  \\
10 & 01 06 30.5 & $-$73 13 58 & 120 \\
11 & 01 11 25.8 & $-$73 14 05 & 120  \\
12 & 00 59 37.3 & $-$73 08 50 & $\ph{1}$90 \\
13 & 01 00 00.0 & $-$72 40 00 & $\ph{1}$60 \\
14 & 00 59 00.0 & $-$72 55 00 & $\ph{1}$90 \\
15 & 01 00 00.0 & $-$72 45 00 & $\ph{1}$90 \\
16 & 00 44 00.0 & $-$73 24 00 & $\ph{1}$94 \\
17 & 00 58 00.0 & $-$72 40 00 & $\ph{1}$90 \\
18 & 01 10 00.0 & $-$72 55 00 & $\ph{1}$86 \\
\hline
\end{tabular}
\end{center}
\end{table}

\subsection{2dF observations}
\label{observations}
The 2dF system is a dual-spectrograph, multi-fibre instrument which
allows up to 400 intermediate-dispersion spectra to be obtained
simultaneously across a two-degree diameter field of view
\citep{lewis02}.  
We used 2dF to observe 18 overlapping SMC fields over 1998
September~25--28 and on 1999 September~30 \& October~1
(Table~\ref{fields}).  The Moon was below the horizon for the majority
of the observations, the exceptions being fields 1, 4, 5, 9, and 10
(Moon before first quarter) and 18 (barely gibbous moon just risen,
$\sim$100$^\circ$ from the SMC).  In a typical observation
approximately 30 `sky' fibres ($\sim$15 for each spectrograph) were
assigned, to ensure good definition of the sky signal.

\begin{figure}
\begin{center}
\includegraphics[width=230pt, angle=0]{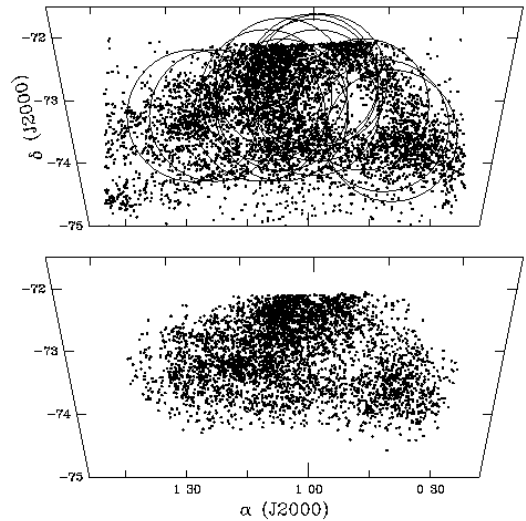}
\caption[]{Spatial distribution of the input catalogue (every 5th
star, upper panel,
observed fields shown as 2$^\circ$ circles) and the
spectroscopically observed targets (all targets, lower panel).}
\label{spatial}
\end{center}
\end{figure}

In order to construct a spectroscopic catalogue that is statistically
representative of the SMC population of hot, luminous stars, targets
were selected from the input catalogue essentially at
random.\footnote{In 1999, some effort was made to redress systematic
under-representation of bright stars in the 1998 run.}  The main
selection criterion was, therefore, simply the set of physical
constraints imposed in configuring 2dF (e.g., avoidance of too-close
fibre heads, large angular deflections of fibres, etc.), rather than
any astrophysical characteristics of the stars beyond those chosen in
constructing the input catalogue.  Usable spectra were obtained for
$\sim$15\%\ of the input catalogue, with a reasonably representative
spatial sampling (Figure~\ref{spatial}).

We used gratings ruled at 1200 lines/mm, which gave coverage of the
$\sim$3900--4800{\AA} region (\Hb\ is included in about half the
spectra) at a resolution of $\sim$2.75{\AA} (FWHM of arc lines; $R
\simeq 1500$), which corresponds to 2.5 pixels on the detectors.
Individual exposures were normally 1800s, and the median continuum
signal-to-noise in the region \lam\lam4395--4460 is $\sim$45, ranging
$\sim$20--150.

Two 2dF datasets were obtained in addition to this main spectroscopic
sample.  First, usable red (H$\alpha$)
spectra were obtained for 1091 targets from the main sample during the
1999 run (September 28 \& 29), again using gratings with 1200 lines/mm
($R \simeq 2500$).

Secondly, in 2001 September, when conducting the programme of
observations of SMC binaries described by \citet*{hhh03}, we were able
to allocate `spare' fibres to targets of interest, with essentially
the same instrumental setup as in 1999.  We chose to observe
UV-selected targets from the UIT catalogue \citep{uitcat97, parker98}
because of their potential interest as an important component of the
hot-star population of the SMC.  Because of the different selection
criteria applied, the 107 targets so observed are separately numbered
in the catalogue: 2dFS\#5001--5107.

\subsection{2dF data reduction}
The bulk of the data reduction was performed using the {\sevensize
2}{\sc dfdr} software \citep{lewis02}.  The main steps include bias
subtraction, extraction of the spectra from the CCD image, optional
division by a normalized flat-field, wavelength calibration,
calibration of the fibre throughputs, and subtraction of the scaled
median sky spectrum.  The software was still undergoing active
development at the time of our reductions, and had been most
extensively tested on datasets quite different to ours (namely, the
2dF quasar and galaxy survey data), so that a number of manual checks
and interventions were necessary.

\subsubsection{Flat-fielding}
At the time of the reductions flat-fielding in {\sevensize 2}{\sc
dfdr} was suboptimal, because any flexure in the spectrographs led to
the extraction of different flat-field pixels compared to those used
in the object frame.  Given the reasonable cosmetic quality of the
detectors (and our familiarity with `known' instrumental features) we
elected not to flat-field our spectra.

\subsubsection{Throughput calibration}
Depending on the configuration, the throughput of individual fibres is
known to vary at the $\sim$10{$\%$} level (separately to the issue of
targets being well centred on the fibre), probably because of
differing stresses within the fibres.  Accurate throughputs are
necessary for reliable subtraction of the sky signal (which is scaled
from sky-only fibres by the relative fibre transmission factors).

At the time of our observations, the normal method of obtaining the
relative throughputs was to take offset-sky exposures in dithered
triplets (to minimize contamination from astrophysical sources)
for each fibre setup.
Observatory recommendations for offset-sky integrations in 1998 (300s)
yielded rather low signal-to-noise for our setup, so longer exposures
were used in 1999 (450s), notwithstanding the larger overheads,
together with twilight throughput frames.  Twilight frames were
acquired in 1998 for only two fields (of course, only the first and
last configurations each night can use these flats for throughputs),
the conventional wisdom at that time (when 2dF observing procedures
were still under development) being that the sky brightness varied
significantly across the 2$^\circ$ field of 2dF.

We found the twilight frames to be entirely consistent with the offset
(dark) skies, with much better signal to noise ratio.  This conclusion
is retrospectively supported by the observations of \citet{ch96}, who
measured the relative brightness gradient of the twilight sky as a
function of angle from the horizon point nearest the Sun.  The 2dF
twilight frames were taken with a maximum zenith distance of
30$^{\circ}$, \ie~the angle between the solar horizon and the observed
region of twilight sky is between 60 and 120$^{\circ}$.  In this range
\citeauthor{ch96} found that the largest observed relative
gradient is $\pm$2{$\%$} per degree of sky, 
which is negligible for our sky-correction purposes.

\subsubsection{Final spectra}

In a number of cases we have repeat observations of a given target
(usually to build up signal-to-noise).  The {\sevensize 2}{\sc dfdr}
algorithm for combining spectra from separate exposures resulted in
nonsensical results in some cases (e.g.,~negative fluxes from the sum
of positive fluxes). {\sevensize 2}{\sc dfdr} has since under{\-}gone
development to remove some of these problems, but we simply summed the
net signals, while looking for (and rejecting) discrepant points; this
proved particularly effective at removing cosmic-ray signatures.

Once combined, the spectra were roughly rectified using a script that
fits a polynomial to predefined continuum regions and then divides the
spectrum through by the fit.  This automated rectification gives
significant time savings and still allows accurate spectral
classification.  Final manual rectifications were conducted as necessary.

\subsection{Long-slit observations}
\label{longslit}

A limited number of supplementary observations 
of 2dF targets were obtained in
conventional long-slit mode, using the Royal Greenwich Observatory
(RGO) Spectrograph at the AAT in July 2001.  Ruled 1200/mm gratings
were used to obtain blue ($\sim$3700--5500\AA) and red
($\sim$5250--7050\AA) spectra.  The mean resolution
element, as defined by the comparison arcs, was 1.6\AA~FWHM,
equivalent to 3.7 pixels.   Standard `optimal' extractions were
performed on these spectra, which 
generally have continuum S/N ratios of
$\sim$100 or better.

The purpose of these observations was not only to obtain repeat
spectra of selected stars of interest, but also to provide a check on
the 2dF data characteristics.  This was considered important in view
of the limited experience of observing `bright', absorption-line
objects with 2dF at the time of our observations, and because the
spectra were extracted with development versions of {\sevensize 2}{\sc
dfdr}.  As discussed in 
\sref{ewerq}, 
the long-slit observations
disclose no obvious problems with the fibre spectra.

\begin{table}
\begin{center}
\caption[]
{Primary temperature-sequence classification criteria applied to the
2dF sample (see Section~\ref{class} for references).  Only some O
subtypes are given here, others being interpolated.}
\label{criteria}
\begin{tabular} {ll}
Type & Criteria \\
\hline
O6 & He~\2 \lam4200 $\sim$ He~\1~{\footnotesize +}~\2 \lam4026 \\
O7 & He~\2 \lam4541 $\sim$ He~\1 \lam4471 \\
O8.5 & He~\2 \lam4541 $\sim$ He~\1 \lam4387 \\
O9 & He~\2 \lam4200 $\sim$ He~\1 \lam4143 \\
\hline
B0 & He~\2 \lam\lam4686, 4541 present, \lam4200 weak \\
B0.5 & He~\2 \lam\lam4200, 4541 absent, \lam4686 weak \\
B1 & He~\2 \lam4686 absent, Si~\4 \lam\lam4088, 4116 present \\
B1.5 & Si~\4 \lam4116 absent, Si~\4 \lam4088 $<$ O~\2 \\
B2 & Si~\4, Si~\2 absent, Si~\3 \lam4553~$>$~Mg~\2 \lam4481 \\
B2.5 & Si~\3 \lam4553 $\sim$ Mg~\2 \lam4481 \\
B3 & Si~\3 \lam4553 $<$ Mg~\2 \lam4481 \\
B5 & Si~\3 absent, Si~\2 \lam4128/4132 $<$ He~\1 \lam4121 \\
B8 & He~\1 \lam4121 $<$ Si~\2 $<$ He~\1 \lam4143 \\
& Mg~\2 \lam4481 $\le$ He~\1 \lam4471 \\
B9 & Mg~\2 \lam4481 $>$ He~\1 \lam4471 \\
& Fe~\2 \lam4233 $<$ Si~\2 \lam 4128/4132 \\
\hline
A0 & Ca$\;K$/(H$\epsilon$ + Ca$\;H$) $<$ 0.33 \\
A2 & 0.33 $<$ Ca$\;K$/(H$\epsilon$ + Ca$\;H$) $<$ 0.53 \\
A3 & 0.53 $<$ Ca$\;K$/(H$\epsilon$ + Ca$\;H$) $<$ 0.75 \\
A5 & 0.75 $<$ Ca$\;K$/(H$\epsilon$ + Ca$\;H$) $<$ 0.85 \\ 
A7 & 0.85 $<$ Ca$\;K$/(H$\epsilon$ + Ca$\;H$) $<$ 0.95 \\
\hline
F0 & Ca$\;K$/(H$\epsilon$ + Ca$\;H$) $\sim$1 \\
F5 & Clear presence of CH $G$-band \\
F8 & $G$-band/H$\gamma$ = 0.5--0.75  \\
G0 & $G$-band/H$\gamma$ = 0.75--0.9  \\
G2 & $G$-band $\sim$ H$\gamma$; H$\gamma$ $>$ Fe~\1 \lam4325 \\
G5 &  H$\gamma$ $\sim$ Fe~\1 \lam4325 \\
G8 &  H$\gamma$ $<$ Fe~\1 \lam4325 \\
\hline
\end{tabular}
\end{center}                        
\end{table}

\section{Spectral Classification}
\label{class}
The 2dF spectra (which consist primarily of B- and A-type stars, with
a minority of O-, F-, and G-types) were classified primarily by visual
inspection, guided by the temperature-sequence classification criteria
summarized in Table~\ref{criteria}.

\subsection{O-type spectra}
\label{otype}
\begin{figure*}
\begin{center}
\includegraphics[height=220mm, width=160mm, angle=0]{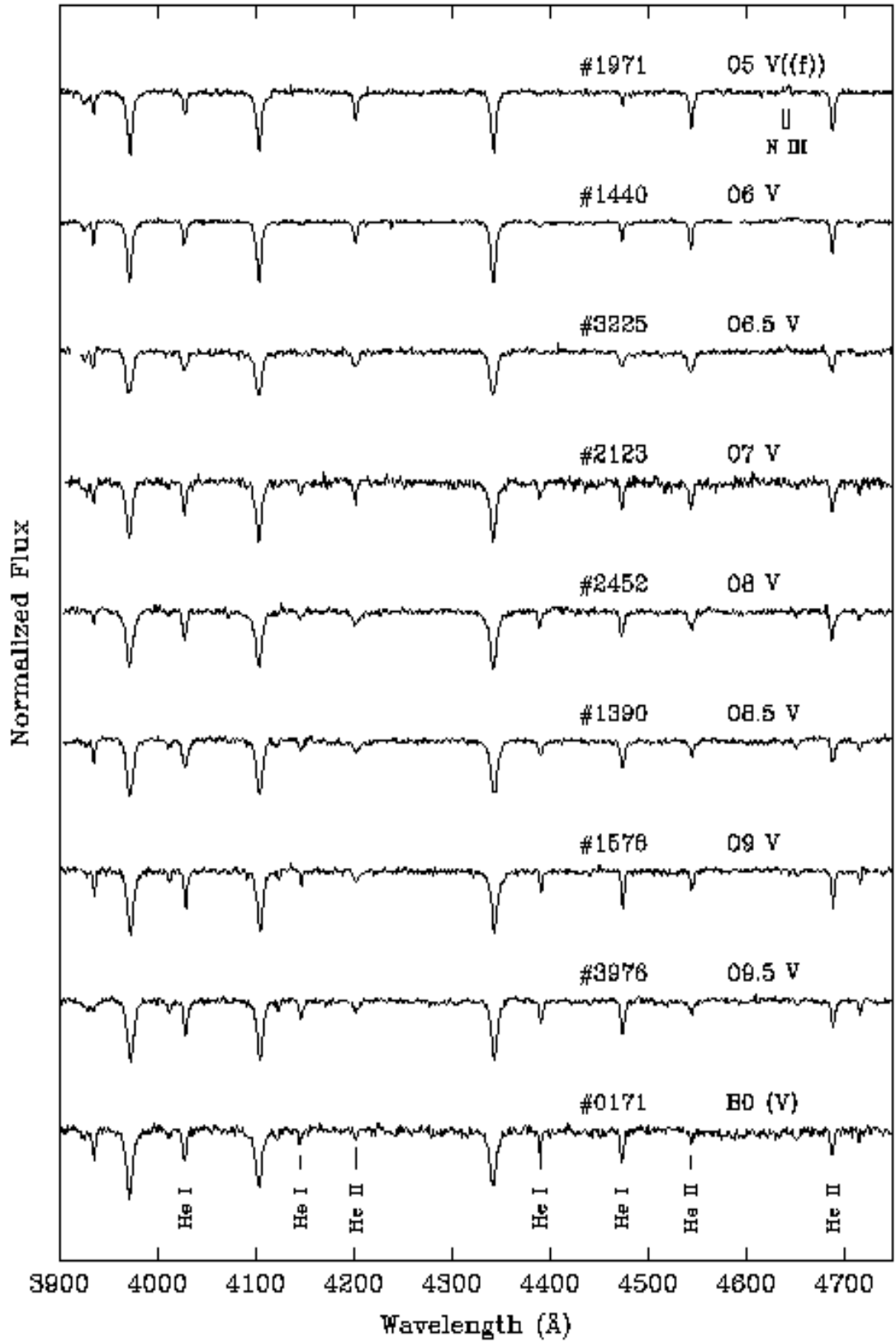}
\caption[] {2dF SMC spectra -- I: O-dwarf spectral sequence. 
Stars are identified by 2dFS catalogue number.  
The luminosity class for the B0 star is assigned using the hybrid
scheme described in Section~\ref{blum}.  The spectral lines
identified in 2dFS\#0171 are, from left to right by species,
He~\1~\lam\lam4026, 4143, 4388, 4471; He~\2 \lam\lam4200, 4541, 4686.
The Balmer lines H$\epsilon$~\lam3970, H$\delta$~\lam4101 and
H$\gamma$~\lam4340 are not explicitly identified.   Note the weak N~\3
\lam4634-40-42 emission and strong He~\2 \lam4686 absorption in 2dFS\#1971,
giving rise to the `((f))' suffix.  Successive spectra
are vertically offset by 0.75 continuum units. }
\label{ostars}
\end{center}
\end{figure*}

Digital spectra are widely available for O-type stars, and the
principal reference we used for the O stars observed with the 2dF was
\citet{wf90}.  The primary classification criteria
are the ratios of He~\1 to He~\2, so the spectral types are largely
unaffected by metallicity effects.  Luminosity classes were assigned
using the precepts given by \citet{wf90}, with reference to the SMC
spectra shown in \citet{wal00}.
Examples of 2dF O-dwarf spectra are shown in Figure~\ref{ostars}.

\begin{figure*}
\begin{center}
\includegraphics[height=220mm, width=160mm, angle=0]{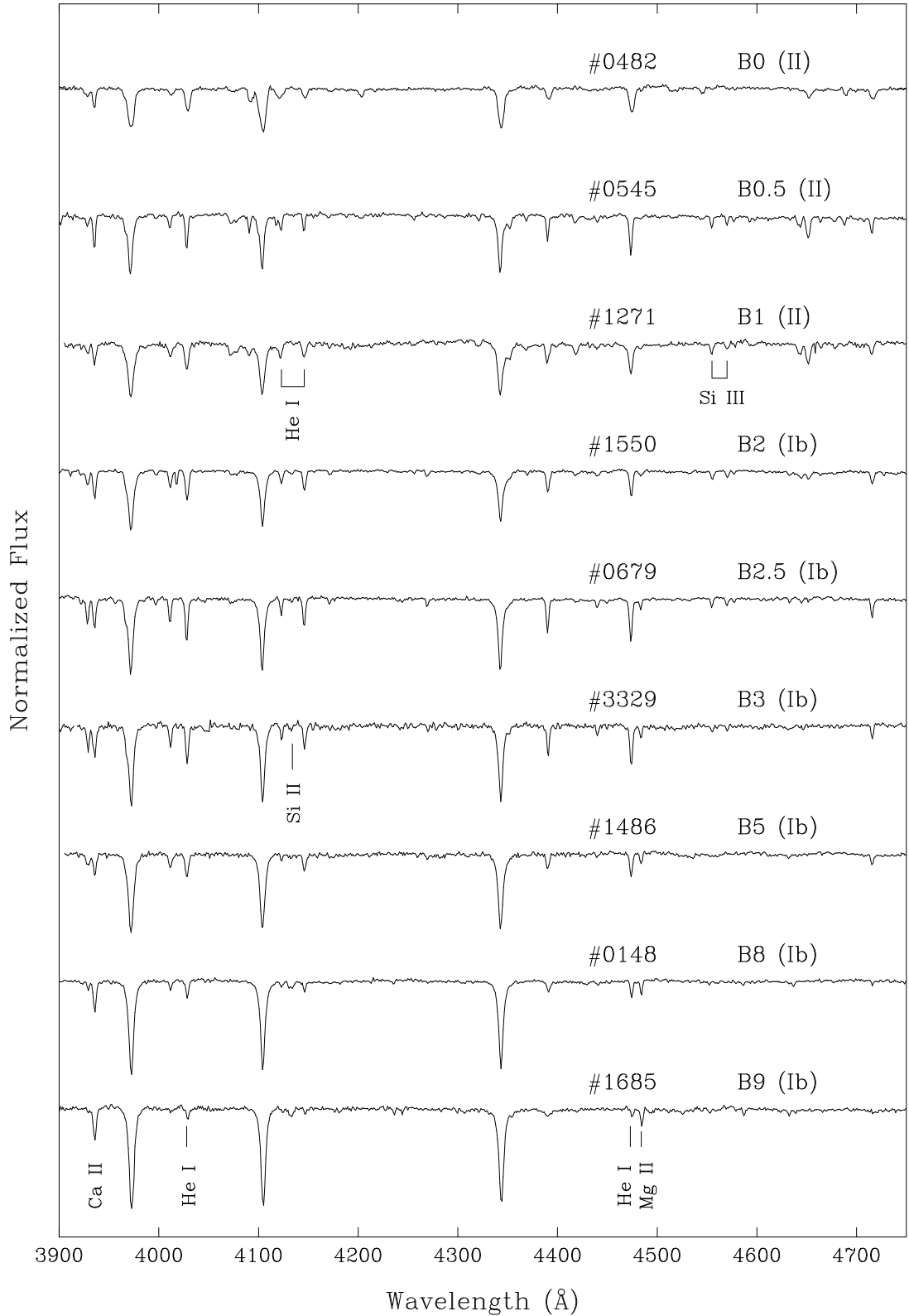}
\caption[] {2dF SMC spectra -- II: B spectral sequence.  
Stars are identified by 2dFS catalogue number.  
The parenthesised luminosity classes indicate their origin in the hybrid
scheme described in Section~\ref{blum}.  The spectral
lines identified in 2dFS\#1685 are, from left to right, Ca~\2
\lam3933 (Ca $K$), He~\1 \lam\lam4026, 4471 and Mg~\2 \lam4481.  The
silicon feature identified in \#3329 is Si~\2 \lam4128/4132 and the
additional lines in \#1271 are He~\1 \lam\lam4121, 4143 and Si~\3
\lam\lam4553, 4568.  Successive
spectra are vertically offset by 0.75 continuum units.
}
\label{bstars}
\end{center}
\end{figure*}

\begin{figure*}
\begin{center}
\includegraphics[height=180mm, width=160mm, angle=0]{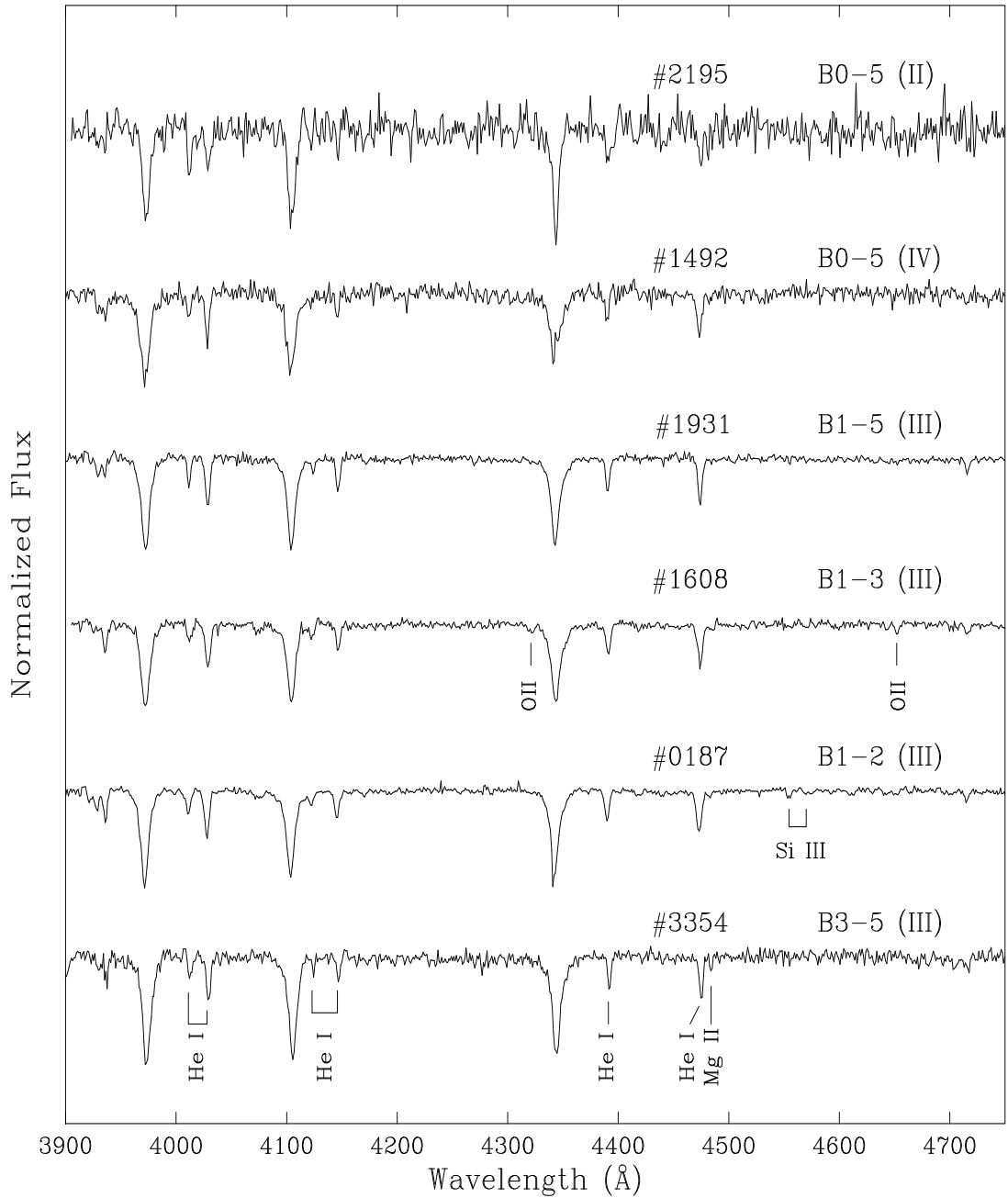}
\caption[] {2dF SMC spectra -- III: Exampes of  B-type spectra without
precise classifications;  cf.\ Section~\ref{btemp}
for details.  Stars are identified by 2dFS catalogue
number.  
The parenthesised luminosity classes indicate their origin in the hybrid
scheme described in Section~\ref{blum}.
The wavelengths of the identified lines are given in
Figs.~\ref{ostars} and~\ref{bstars} with the addition of He~\1
\lam4009 in 2dFS\#3354 and the O~\2 \lam4317--19 and 4650 blends
in \#1608.  Successive spectra are vertically offset by 0.75 continuum units.
}
\label{bstars2}
\end{center}
\end{figure*}

\begin{figure*}
\begin{center}
\includegraphics[height=180mm, width=160mm, angle=0]{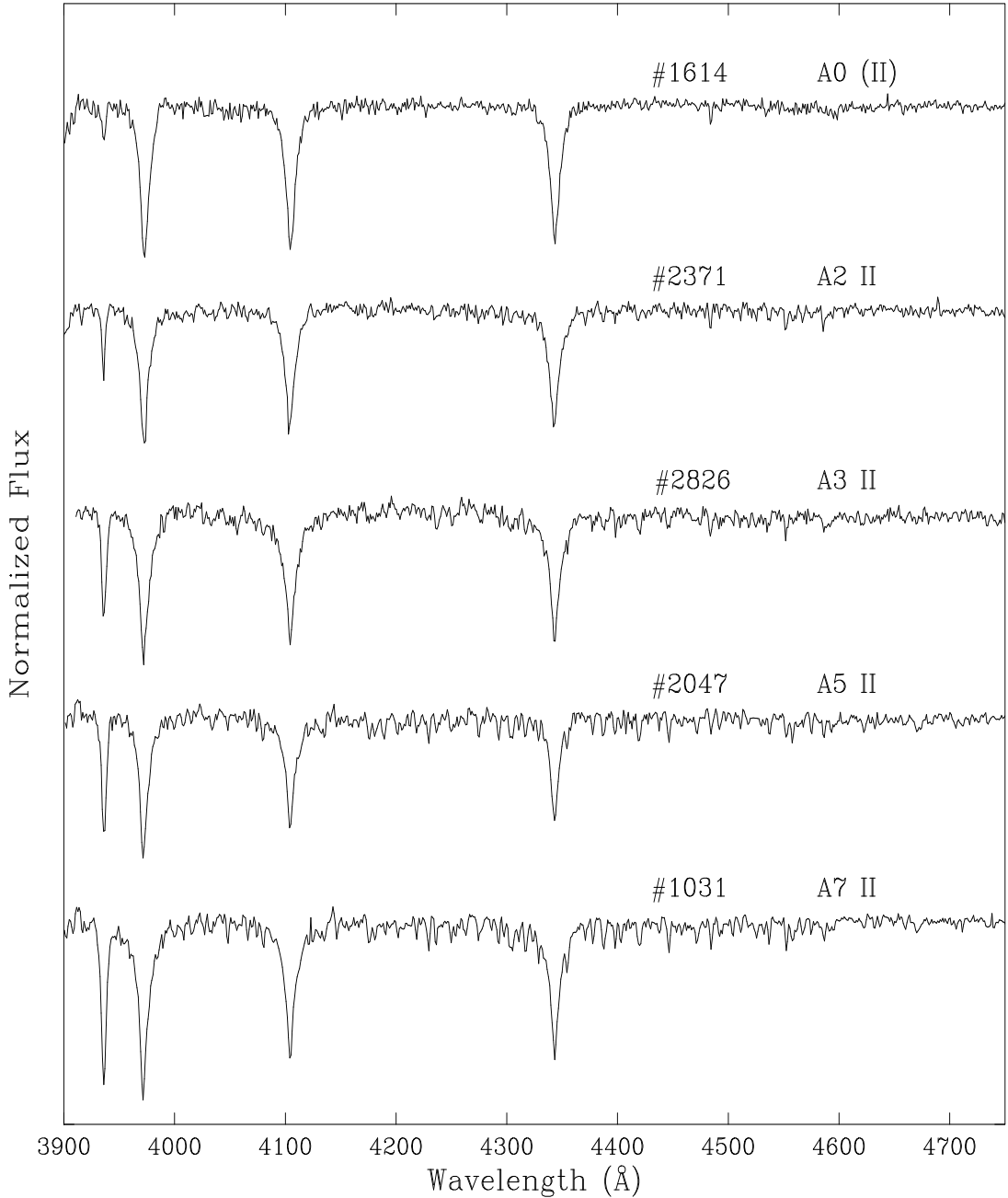}
\caption[]
{2dF SMC spectra -- IV: Exampes of A-type spectra.  Stars are
identified by 2dFS catalogue number.  
The luminosity class for the A0 star is assigned using the hybrid
scheme described in Section~\ref{blum}.  Successive spectra are
vertically offset by one continuum unit.  }
\label{astars}
\end{center}
\end{figure*}

\subsection{B-type spectra}
\label{btype}

\subsubsection{Temperature sequence}
\label{btemp}
For the B-type stars the principal reference we adopted was Lennon's
(1997) study of SMC B-type supergiants; the primary classification
criteria used here are included in Table~\ref{criteria}, and,
excepting the very earliest subtypes, use metal lines.

The overall strength of the metallic lines in B-type spectra is
related to the luminosity: the more luminous the star, the stronger
the lines \citep[e.g.,][]{wf90}.  The majority of the 2dF targets are
substantially less luminous than those in Lennon's sample; thus the
metal lines in the 2dF B-star spectra are weak both because of the
reduced metallicity of the SMC, and also because of their relatively
low luminosity.  As a consequence, while some of the 2dF spectra can
be precisely classified following Lennon's scheme, the large majority
require coarser classification bins, even where the data quality is
moderately good.

By way of illustration, consider a mid-B giant where the Si~\3
\lam4553 line is undetectable.  This may be because it is B5 or later,
or because it is slightly hotter (e.g.,~B3) but, due to
luminosity effects, the Si~\3 line is weak and in the noise.
These problems mainly arise in the B1--B5 range, where the
primary criteria involve the ratios of Mg~\2 \lam4481 and
Si~\3.  Inspection of Lennon's data also reveals that the O~\2
\lam4415--17, 4640-50 and N~\2 \lam4631 features are
undetectable at B3 and later; when observed, these features 
therefore allow the spectral type to be more closely bracketed, to~B1--B3.

In Lennon's scheme, type B1 is characterized by weak or absent He~\2
\lam4686; however, the 2dF spectra are at lower resolution than his
data ($\sim$2.5--3~vs~1.2\AA), and poorer signal-to-noise.  The framework
adopted here uses the intermediate B0.5 type for those stars where
He~\2 \lam\lam4200, 4541 are absent and weak \lam4686 is seen.

Examples of precisely classified 2dF B-type spectra are shown in
Figure~\ref{bstars}; further examples, where precise classification
was not possible (even at fairly good S/N), are shown in
Figure~\ref{bstars2}.  The spectrum of 2dFS\#2195 is included
in Figure~\ref{bstars2} as an example of the low signal-to-noise 2dF
data; the star is clearly of early-B type (given the absence of strong
He~\2 lines and the strength of the He~\1 lines), but cannot be more
accurately classified.  A similar classification is applied to the
spectrum of 2dFS\#1492, even though the data are of substantially better
quality.  It is not possible to be confident that the weak He~\2 lines
seen in, e.g.,~2dFS\#0482 are absent in \#1492, and the
metal lines necessary for unique classification in the early B-stars
are not seen (as in 2dFS\#1931).

Many of the B-type spectra display emission in one or more of the
Balmer lines (see Section~\ref{Bem_stat}).  Where double-peaked or
resolved Balmer emission is evident, an `e' qualifier is adopted (see
Section~\ref{Bem_txt}).

\subsubsection{Luminosity classification}
\label{blum}

In principle, luminosity-class assignments can be made using
combinations of features, as for Galactic stars, modified to account
for the systematic differences in metal-line strengths (e.g.,
\citealt{wal77}).
We attempted to classify the B-type stars in our sample using
this approach, but were handicapped by the generally inadequate
signal-to-noise of our data.  We were therefore forced to fall back on
the use of the Balmer lines alone, and particularly \Hg.
Notwithstanding potential pitfalls (cf.\ \citealt{jj}), this provides
the simplest criterion for luminosity classification of B (and A)
stars.

We measured the \Hg\ equivalent widths, \Wl, in the entire 2dF sample
`by hand', and by using an automated gaussian-fitting procedure.  In
addition to \Wl, the gaussian fits yield the actual line width,
characterized by full width at half depth.  Excepting a small minority
of pathological cases (e.g., stars with strong line-core emission,
where the automated fits give erratic results), all three measures
lead to essentially identical conclusions in respect of
classifications; in particular, the automated and `by hand' equivalent
widths are in excellent agreement.  For reasons of objectivity and
reproducibility, we generally utilize the gaussian-fit equivalent
widths.

\begin{table}
\caption{B-star luminosity-class criteria:  boundary values
of the parameter $B+0.3\Wl$ (cf.\ Section~\ref{blum}).  Parentheses
are used to indicate that these are not true
morphological luminosity classes, but are based on both 
spectroscopic {\em and} photometric criteria.}
\begin{tabular}{lcccccccccccccc}
\hline
\multicolumn{1}{|c|}{Sp.}&
\multicolumn{14}{|c|}{Luminosity Class}\\
\multicolumn{1}{|c|}{Type}&
\multicolumn{2}{|c|}{(Ia)} &  
\multicolumn{2}{|c|}{(Iab)}& 
\multicolumn{2}{|c|}{(Ib)} & 
\multicolumn{2}{|c|}{(II)} & 
\multicolumn{2}{|c|}{(III)}& 
\multicolumn{2}{|c|}{(IV)} & 
\multicolumn{2}{|c|}{(V)}  \\
\hline
\multicolumn{2}{l}{B0}     &\multicolumn{2}{|c|}{12.27}&\multicolumn{2}{|c|}{12.92}&\multicolumn{2}{|c|}{13.35}&
          \multicolumn{2}{|c|}{13.96}&\multicolumn{2}{|c|}{14.80}&\multicolumn{2}{|c|}{15.57}\\
\multicolumn{2}{l}{B0.5}   &\multicolumn{2}{|c|}{12.30}&\multicolumn{2}{|c|}{12.95}&\multicolumn{2}{|c|}{13.82}&
          \multicolumn{2}{|c|}{14.43}&\multicolumn{2}{|c|}{14.82}&\multicolumn{2}{|c|}{15.60}\\
\multicolumn{2}{l}{B1}     &\multicolumn{2}{|c|}{12.34}&\multicolumn{2}{|c|}{13.12}&\multicolumn{2}{|c|}{14.00}&
          \multicolumn{2}{|c|}{14.92}&\multicolumn{2}{|c|}{15.76}&\multicolumn{2}{|c|}{16.47}\\
\multicolumn{2}{l}{B1.5}   &\multicolumn{2}{|c|}{12.09}&\multicolumn{2}{|c|}{13.14}&\multicolumn{2}{|c|}{14.09}&
          \multicolumn{2}{|c|}{15.27}&\multicolumn{2}{|c|}{16.35}&\multicolumn{2}{|c|}{17.07}\\
\multicolumn{2}{l}{B2}     &\multicolumn{2}{|c|}{12.10}&\multicolumn{2}{|c|}{13.15}&\multicolumn{2}{|c|}{14.23}&
          \multicolumn{2}{|c|}{15.54}&\multicolumn{2}{|c|}{16.63}&\multicolumn{2}{|c|}{17.41}\\
\multicolumn{2}{l}{B2.5}   &\multicolumn{2}{|c|}{12.16}&\multicolumn{2}{|c|}{13.20}&\multicolumn{2}{|c|}{14.28}&
          \multicolumn{2}{|c|}{15.73}&\multicolumn{2}{|c|}{16.81}&\multicolumn{2}{|c|}{17.72}\\
\multicolumn{2}{l}{B3}     &\multicolumn{2}{|c|}{12.48}&\multicolumn{2}{|c|}{13.39}&\multicolumn{2}{|c|}{14.40}&
          \multicolumn{2}{|c|}{16.18}&\multicolumn{2}{|c|}{17.72}&\multicolumn{2}{|c|}{18.63}\\
\multicolumn{2}{l}{B5}     &\multicolumn{2}{|c|}{12.71}&\multicolumn{2}{|c|}{13.49}&\multicolumn{2}{|c|}{14.69}&
          \multicolumn{2}{|c|}{16.99}&\multicolumn{2}{|c|}{18.86}&\multicolumn{2}{|c|}{19.51}\\
\multicolumn{2}{l}{B8}     &\multicolumn{2}{|c|}{13.11}&\multicolumn{2}{|c|}{13.88}&\multicolumn{2}{|c|}{15.26}&
          \multicolumn{2}{|c|}{18.19}&\multicolumn{2}{|c|}{20.54}&\multicolumn{2}{|c|}{21.18}\\
\multicolumn{2}{l}{B9}     &\multicolumn{2}{|c|}{13.22}&\multicolumn{2}{|c|}{14.37}&\multicolumn{2}{|c|}{15.96}&
           \multicolumn{2}{|c|}{18.91}&\multicolumn{2}{|c|}{21.37}&\multicolumn{2}{|c|}{21.90}\\
\multicolumn{2}{l}{A0}     &\multicolumn{2}{|c|}{13.37}&\multicolumn{2}{|c|}{14.60}&\multicolumn{2}{|c|}{16.86}&
           \multicolumn{2}{|c|}{20.10}&\multicolumn{2}{|c|}{22.16}&\multicolumn{2}{|c|}{22.76}\\
\hline
&&&&&&&&&&&&&&\\
\end{tabular}
\label{blumtab}
\end{table}

Ideally, our luminosity-class assignments should be based exclusively
on spectral morphology.  Unfortunately, the spread in \Hg\ equivalent
width is fairly large at given $B$ magnitude (as discussed further in
Section~\ref{MBWL}); equivalently, the range in $M(B)$ at given
\Wl\ is considerable.  Thus our \Wl\ data, which are not particularly
accurate because of the limited signal-to-noise ratio in the spectra, are
of rather limited utility in predicting absolute magnitude and, by
inference, luminosity class, particularly at early-B types, where
stars with the same \Wl\ are found at all luminosity classes (cf.\
Fig~\ref{EWB_B}; B0$\;$V and B0$\;$I stars differ in \Wl\ by only
$\sim$2\AA, comparable with the spread in the measurements).  Faced
with this difficulty, and after considerable experimentation, we
reluctantly abandoned the principled position of luminosity
classification based solely on morphology in favour of a more
pragmatic approach incorporating absolute-magnitude information.

The (ad hoc) formulation we adopted is that the boundaries between
luminosity classes are defined by loci of constant $B+0.3\Wl$, as
listed in Table~\ref{blumtab}.  
To define the boundaries, we adopted the
relationship between $M(V)$ and luminosity class given by
\citet{hm84}, augmented by \citet{sk82} as necessary, and converted to
$B$ by adopting $(B-V)_0$ colours from
\citet{fg70} and an {\em apparent} distance modulus of 19.2, from
\citet{hhh03}.  The equivalent width corresponding to each $B$
magnitude was then determined directly from the data (cf.\
Section~\ref{MBWL}).

Clearly, the resulting `luminosity classes' are not true, MK-process,
morphological types, and to make this clear the luminosity-class
assignments derived through the hybrid photometric/spectroscopic
approach are given in parentheses in the catalogue, and throughout
this paper.  Nonetheless, the hybrid classifications do appear to be
broadly consistent with results from other sources
(Section~\ref{sp_comp}), although for spectra where \Hg\ is filled in
by emission our assigned luminosity classes are liable to be too
bright.  For stars of particular interest, investigators should
therefore refer to the notes on Balmer emission that we provide for
every star (Section~\ref{Bem_stat}; Appendix~A).

\subsection{A-type stars}
\label{atype}

\subsubsection{Temperature sequence}

The A-type stars were classified on the basis of the
Ca~$K$/H$\epsilon$ line ratio, as discussed by \citet{eh03}.  Reliance
on the calcium line ratios is not without its problems (see discussion
in \citeauthor{eh03}), but it is quick to apply and does not rely on weak
metal lines.  The classification criteria from \citeauthor{eh03} are
included in Table~\ref{criteria}.  Figure~\ref{astars} illustrates
a spectral sequence, and Section~\ref{AFcomp} discusses a handful of
peculiar A~stars.

\begin{table}
\begin{center}
\caption{Galactic A-type stars observed for calibration purposes.
Most of these are discussed by \citet{eh03}; new stars are identified
by lower-case spectral-type sources.}
\label{GalA}
\begin{tabular}{rlccrlc}
\hline
HD$\phantom{4}$&
\multicolumn{1}{c}{Spectral}&
Source&
&
HD$\phantom{4}$& 
\multicolumn{1}{c}{Spectral}&
Source\\
&
\multicolumn{1}{c}{type}&      
&
&  
& 
\multicolumn{1}{c}{type}&  \\
\hline 
      1404  &  A2 V        &GG7&\quad&     175687  &  A0 II       &c69   \\
      1457  &  A9 Ib--II   &g01&\quad&     186177  &  A5 II       &g01   \\
      3283  &  A3 II       &GG9&\quad&     187983  &  A1 Iab      &M55   \\ 
      3940  &  A1 Ia       &M55&\quad&     192514  &  A3 III      &S54   \\
      8538  &  A5 V        &M53&\quad&     195324  &  A1 Ib       &C69   \\
     10845  &  A8 III      &GG9&\quad&     196379  &  A9 II       &M73   \\
     12216  &  A1 V        &S54&\quad&     197345  &  A2 Ia       &M73   \\
     12279  &  A0 V        &GG7&\quad&     197489  &  A5 II       &g01   \\
     12953  &  A1 Ia       &M55&\quad&     201935  &  A5 II       &g01   \\
     13041  &  A4 V        &GG9&\quad&     202240  &  A8 II       &g01   \\
     13476  &  A3 Iab      &M55&\quad&     203280  &  A7 IV--V    &M53   \\
     14433  &  A1 Ia       &m55&\quad&     205835  &  A5 V        &GG9   \\
     14489  &  A2 Ia       &M55&\quad&     207260  &  A2 Ia       &M55   \\
     14535  &  A2 Ia p:    &m55&\quad&     207673  &  A2 Ib       &M55   \\
     15316  &  A3 Iab      &M55&\quad&     210221  &  A3 Ib       &M55   \\
     17378  &  A5 Ia       &M55&\quad&     211868  &  A5 Ib--II   &g01   \\
    148743  &  A7 Ib       &G01&\quad&     212511  &  A3 II       &g01   \\
    161695  &  A0 Ib       &C69&\quad&     213558  &  A2 V        &S54   \\
    164514  &  A5 Ia       &m55&\quad&     213973  &  A9 III      &A81   \\
    165784  &  A2 Ia       &m55&\quad&     216701  &  A7 IV       &GG9   \\
    167356  &  A0 Ia       &m55&\quad&     220770  &  A5 Ib       &m55   \\
    172167  &  A0 V        &M73&\quad&     222275  &  A5 III      &A85   \\
    173880  &  A3 V        &GG9&\quad&     223385  &  A3 Ia       &M55   \\
\hline
\end{tabular}
\end{center}
Sources of spectral types are (in order of preference):
M73, \citet{mk73};
M43, \citet{mkk};
M55, \citet{mcw55};
M53, \citet{mhj53};
M50, \citet{mr50};
S54, \citet{slet54};
GG7, \citet{gg87};
GG9, \citet{gg89b};
A81, \citet{abt81};
A85, \citet{abt85};
G01, \citet{gnw01};
C69, \citet{ccjj}.
\end{table}

\begin{table}
\begin{center}
\caption{Adopted luminosity-class criteria:  
luminosity class for A2--A7 stars as a function
of \Hg\ equivalent width, in~\AA\ (cf.~Section~\ref{gal_cal}).
\label{HGlum}}
\begin{tabular}{ccccc}
\hline
Ia   & Iab  & Ib   & II    & III \\
$<3$ & 3--4 & 4--5 & 5--10 & 10--15 \\
\hline
\end{tabular}
\end{center}
\end{table}

\subsubsection{Luminosity classification}
\label{gal_cal}

\citet{eh03} describe observations of a number of Galactic BAF stars
acquired for comparison and calibration purposes.  We have
supplemented those observations with some new data, giving a total of
46 A-type stars (Table~\ref{GalA}).  We measured \Hg\ in these
Galactic standards in the same way as in our 2dF sample, in order to
calibrate line strength and luminosity class as a function of spectral
subtype.  All the available data are consistent with the simple
luminosity-class--equivalent-width scheme summarized in
Table~\ref{HGlum}, which is independent of A~spectral subtype for the
observed Galactic standards.

The adopted luminosity-class bins are broadly consistent with the
calibrations given by \citet{bc74} and \citet{azzo87}, except that
\citeauthor{azzo87}'s equivalent widths are 1--2\AA\ ($\sim$30--50\%) larger
than ours for late-A supergiants.  This discrepancy appears to arise
because \citeauthor{azzo87}'s (mainly interpolated) late-A values are
influenced by \Hg\ equivalent widths for F-type stars that are larger
than we find (Section~\ref{mbwlover}).  Although late-A, bright
supergiants are not well represented in either sample, our dataset is
more extensive, and of better quality, than his.

In our SMC dataset, the \Hg\ equivalent widths at given $B$
magnitude are smaller for A0 stars than for later A
types.  We applied the Galactic calibration to A2--A7 subtypes
(thereby assigning supergiant luminosity classes to some rather faint
stars -- see Fig.~\ref{EWB_A}), while treating A0 stars in the same
way as B-type stars (Table~\ref{blumtab}).

\begin{figure*}
\begin{center}
\includegraphics[height=198mm, width=160mm, angle=0]{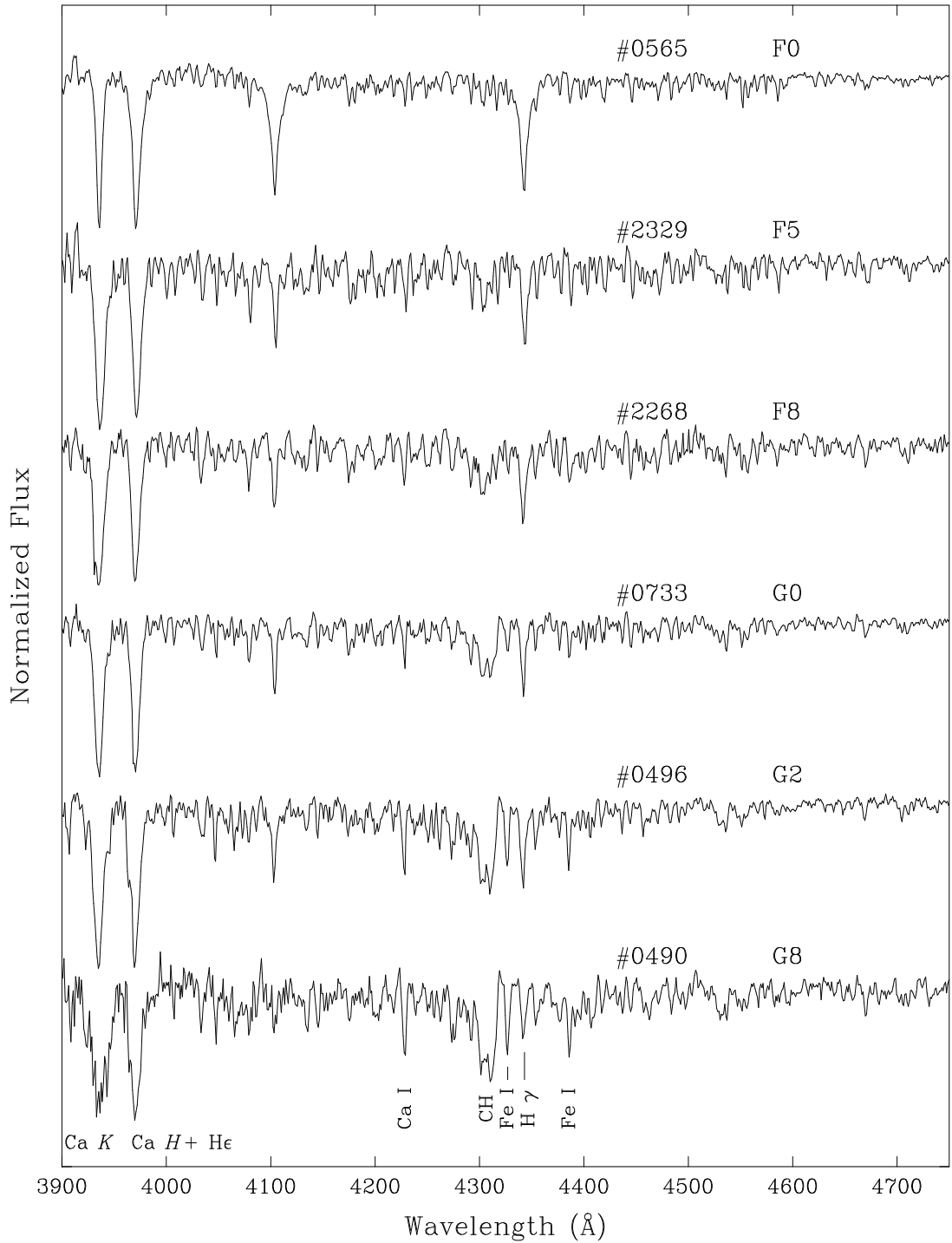}
\caption[] {2dF SMC spectra -- V: F- and G-star spectral sequence.  
Stars are identified by 2dFS catalogue number.  The spectral lines
identified in 2dFS\#0490 are 
Ca~$K$ (\lam3933), the  Ca $H$ (\lam3969)
\& H$\epsilon$ blend,
Ca~\1~\lam4226, the CH $G$-band (around
\lam4300), Fe~\1 \lam\lam4325, 4383, and H$\gamma$.  
Successive
spectra are vertically offset by one continuum unit.}
\label{fstars}
\end{center}
\end{figure*}

\subsection{Later types}

The remaining `late'-type spectra with radial velocities consistent
with SMC membership were classified using the criteria
listed by 
\citet{eh03}.
Briefly, F- and G-type spectra are
characterized by increasing metal-line strength as the
temperature decreases; for completeness, the criteria are included
here in Table~\ref{criteria}.  A spectral sequence is shown in
Figure~\ref{fstars}.

Since our work is primarily concerned with the OBA domain, we have not
explored luminosity-class diagnostics in later-type stars
in the sample, though from
their physical luminosities these objects are  expected to
be luminosity classes Ib--II for the most part.

\section{Classification checks}
\label{chx_sx}
\begin{table}
\begin{center}
\caption[]
{Comparisons of 2dF spectral types with those by 
\citet[][ first nine entries]{djl97}
and \citet{pm95}.   Parenthesised 2dF luminosity classes were assigned
using the hybrid scheme described in Section~\ref{blum}.}
\label{LenMas}
\begin{tabular}{ccll}
\hline
2dFS \# & AzV \# & 2dF & Other \\
\hline
0577 & {\pA}10 & B2.5 (Iab) & B2.5 Ia \\
0801 & {\pA}96 & B1.5 (Iab) & B1.5 Ia \\
1352 & 215     & B0 (Ib)    & BN0 Ia \\
1550 & 268     & B2 (Ib)    & B2.5 Iab \\
2174 & 404     & B2.5 (Iab) & B2.5 Iab \\
2538 & 445     & B5 (Iab)   & B5 Iab \\
2773 & 462     & B2 (Iab)   & B1.5 Ia \\
2907 & 472     & B2 (Iab)   & B2 Ia \\
3235 & 490     & O9.7 Ia+   & O9.5 II \\
\hline
0668 & {\pA}28 & B1-2 (II)  & B1 I \\
0764 & {\pA}73 & B0 (III)   & O8.5 V  \\
0786 & {\pA}84 & B0.5 (IV)  & B1 V \\ 
0836 & 114     & O8 V       & O7.5 V \\
1271 & 196     & B1 (II)    & B0.5 III \\
1324 & 209     & B0.5 (IV)  & B1 V \\
1357 & 217     & B1-3 (II)  & B1 III \\
1527 & 261     & B2 (Ib)e   & O8.5 I  \\
1545 & 266     & B1 (Iab)   & B1 III \\
1550 & 268     & B2 (Ib)    & B2.5 V \\
1654 & 302     & O9 III     & O8.5 V \\
1741 & 326     & B0 (IV)    & O9 V \\ 
1759 & 328     & B0 (III)   & B0 V \\
1766 & 334     & O9.5 III   & O8.5 V \\
1858 & 346     & B1-5 (II)  & B2 V \\
1879 & 351     & B1-2 (Ib)e & B0 V \\
1904 & 354     & B1-3 (II)  & B1.5 V \\
1972 & 376     & B1-2 (Iab) & Be \\ 
2033 & 386     & B1-2 (II)  & B1.5 V \\ 
2102 & 395     & B3 (II)e   & B1 III \\
2139 & 402     & O9.7 Iab   & Be \\
2201 & 409     & B0.5 (Ib)  & Be \\
2319 & 423     & B0 (II)    & O9.5 V \\
2413 & 436     & B0 (II)e   & O7.5 Ve \\
2717 & 456     & O9.5 Ib    & O9.5 V \\
2720 & 457     & B1-3 (II)e & Be \\ 
2905 & 471     & B0 (III)   & B0.5 V \\
3047 & 480     & O4-7 Ve    & Oe \\
3249 & 491     & B1-3 (III) & O7 III: \\
3530 & 503     & B0-5 (II)e & Be \\
\hline
\end{tabular}
\end{center}
\end{table}

\subsection{Comparison with other sources}
\label{sp_comp}

Nine of the stars in the 2dF sample were observed by \citet{djl97}.
Lennon's are high-quality classifications, and provide an external
check on the current work.  Table~\ref{LenMas} shows the 2dF types
compared to Lennon's classifications.  With the exception of AzV~490,
both the spectral and luminosity types agree to within one subtype;
the differences for AzV~490 are astrophysical (Section~\ref{azv490}).

\begin{figure*}
\begin{center}
\includegraphics[width=125mm, height=160mm, angle=270]{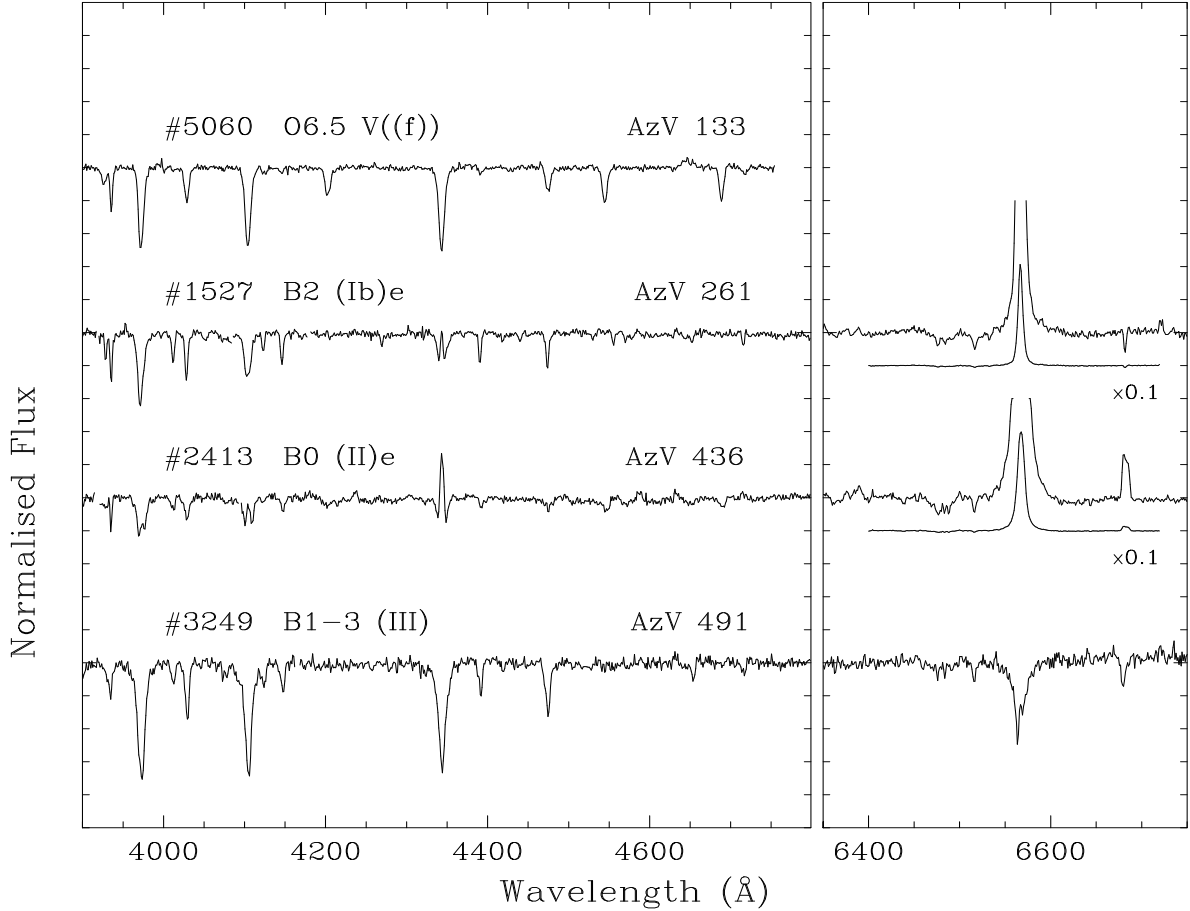}
\caption[ ] {2dF spectra for previously classified
stars in the AzV catalogue (see
\sref{sp_comp} for details). Spectra are offset by 0.5 continuum units.}
\label{spectra4}
\end{center}
\end{figure*}

Table~\ref{LenMas} also compares the 2dF spectral types with those
given by \citet{pm95} for 30 stars in common.  The differences in
spectral types are again generally not larger than one subtype,
exceptions being AzV~261, 436 and 491, where our types are
significantly later.  Re-examining the 2dF spectra of these stars
(Fig.~\ref{spectra4}), we could not justify revising our results.
Neither AzV~261 nor 491 shows He~\2 in our data, and, while the two
2dF spectra of AzV~436 at our disposal have low signal-to-noise, He~\1
\lam4144 appears stronger than He~\2~\lam4200 in both, implying a
type later than O9.

\citet{pm02} presents a compilation of spectral classifications 
for 436 stars, drawn from a variety of sources (but omitting Lennon's
work); of these, 230 have slit spectra, the remainder being
objective-prism results.  We have 2dF classifications for 158 stars
from the full listing; comparison with the long-slit types shows
agreement to within, normally, 1--2 subtypes.  There are two notable
exceptions, namely AzV~133 and 336
(2dFS\#5060 and 1776).  \citet{gar87} classified AzV~133
as B0n$+$O8: with the comment that broad, possibly double lines of the
Balmer series were seen; in the 2dF data 
(Fig.~\ref{spectra4}) the Balmer lines appear
unremarkable, and we classify the spectrum as O6.5~V((f)).
\citet{pm02} gives the spectral type for AzV~336 as WN2$+$abs
(cf. B2~(III) from the 2dF spectrum); the WN star is actually 
the separate object AzV~336A
\citep[see][]{pm01}.

\subsection{Multiple Observations}

Fifty targets were observed with 2dF in both 1998 and 1999.  These
spectra were classified entirely independently, and test the internal
consistency of the classifications.  In all instances the spectral
types from the two seasons agree to within one spectral subtype at the
appropriate classification resolution.  The quality of the data is
such that small differences are not unexpected, especially in cases
where the line ratios are near the edges of the classification bins.

The long-slit spectra from 2001 provide a further consistency check.
The spectral types again agree with the 2dF results to within one
spectral subtype.

Overall, therefore, we believe that, notwithstanding the 
limitations of our data, in general our classifications are consistent
to within a spectral subtype at the (sometimes coarse) classification
bins adopted.

\begin{figure*} 
\begin{center}
\includegraphics[width=420pt]{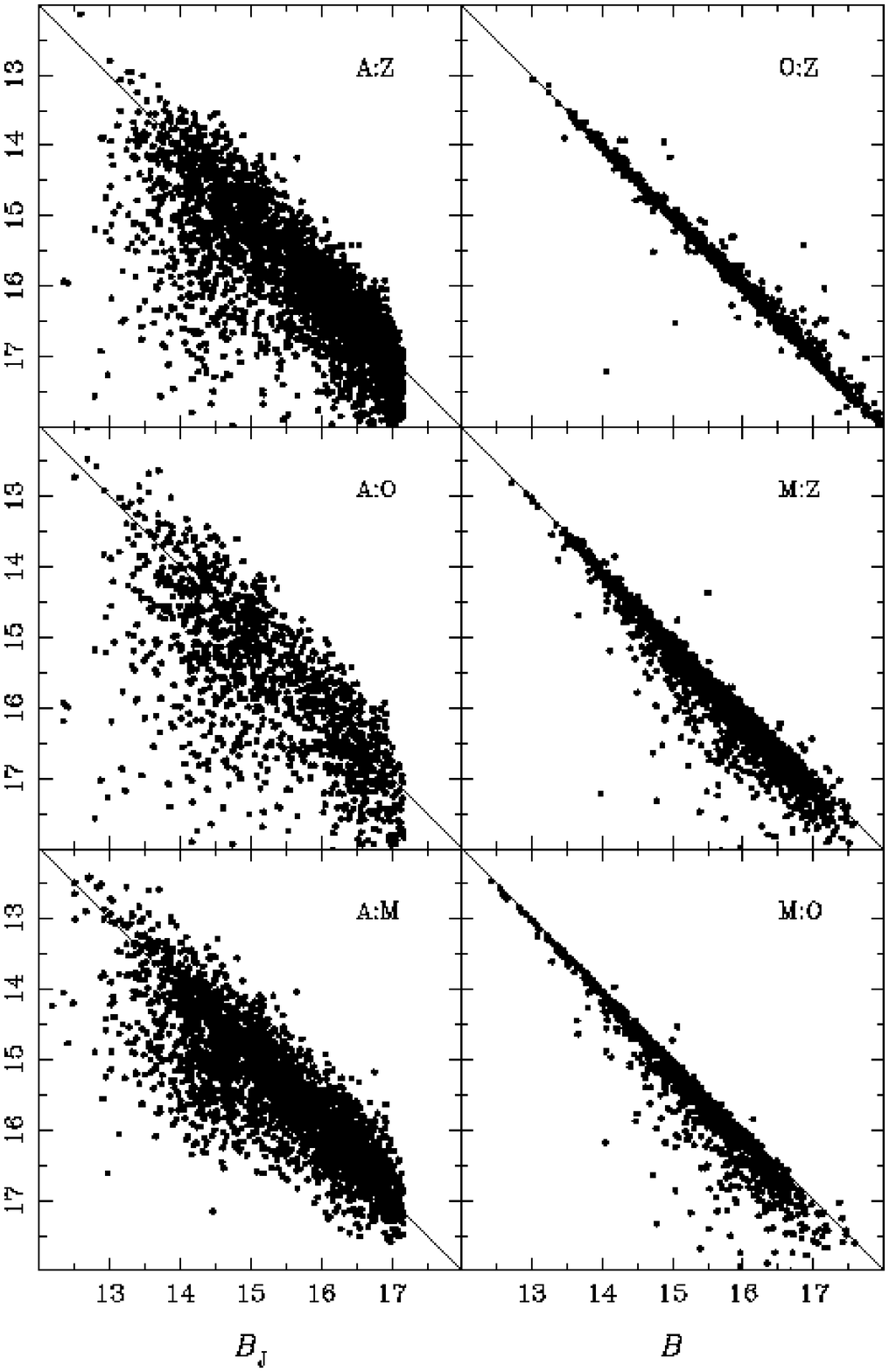}
\caption{Comparisons of photometric results for stars in the 2dF
survey, from APM, \citeauthor{pm02}, MCPS and OGLE (labelled A, M, Z and O,
respectively, in this figure, with the $x$ axis dataset listed first).
\label{Phot1}}
\end{center}
\end{figure*}

\section{Photo{\-}metry and astrometry}
\label{sec_phot}

Initial construction of the 2dF input catalogue was necessarily
performed using APM photographic photo{\-}metry for colour--magnitude
selection.  Subsequently, three important catalogues of CCD-based
photo{\-}metry of stars in the SMC have been published: OGLE
\citep{u98}, MCPS \citep{z02} and \citet{pm02}.   
We have used these catalogues to refine and extend photo{\-}metry for our
targets, and to characterize the input material.

We sought positional matches between the 2dF catalogue and the
photometric catalogues, initially by coincidence within a circle of
radius 3$''$.  We then adjusted the catalogue co-ordinates by the
(subarcsecond) median offsets in RA and declination, and then
sequentially reduced the adopted `match' radius while further
adjusting the offset as necessary.  (Using the median rather than the
mean offset ensured that this process converged in zero or one
iteration.)

This comparison is limited to the main catalogue of 4054 2dF targets,
as we have incomplete APM data for the supplementary UIT targets,
for which OGLE astrometry was used for spectrograph configuration
(cf.~\citealt{hhh03}).

\subsection{OGLE}

The final adopted `coincidence' radius was 1.2$''$, generating 1395
matches, plus a further 290 duplicates (cp.\ 1461, 9108 at 5$''$).
The duplicates were resolved by selecting the OGLE star closest in $B$
magnitude to APM \BJ\ (after applying the small zero-point correction
to \BJ\ discussed below).  This proved very effective; spot checks of
a few dozen duplicates showed that in every case there was one
`obvious' candidate (invariably the brightest OGLE star in the
coincidence radius).

\subsection{MCPS}

The same procedures were adopted for the MCPS, with a 1.2$''$ search
radius yielding 3231 matches plus 67 duplicates, again resolved by
magnitude matching.  (Increasing to 5$''$ adds a further 158 matches
and 5797 duplicates.)  This includes 15 stars where \citeauthor{pm02}
photo{\-}metry has been adopted for the MCPS (see \citealt{z02}).

\subsection{Massey}

\citeauthor{pm02}'s survey was aimed at obtaining 
photo{\-}metry for the brighter MC stars in particular.  It has,
accordingly, a brighter magnitude limit, allowing a larger plate scale
than the other CCD-based surveys (2.3$''$/pixel vs.\
$\sim{0.7}''$/pixel).  Nonetheless,
\citeauthor{pm02}'s astrometry is sufficiently good that a 1.5$''$
coincidence radius proved satisfactory, yielding 3115 matches, plus 47
duplicates (cp.\ 3211, 90 at 5$''$).

We found that all the duplicates appear to be double entries in 
\citeauthor{pm02}'s catalogue; we simply adopted results for the first
entry. Further examination of this catalogue shows 4379 cases where
two or more entries are spatially coincident to within 2.3$''$ (3503
pairs with $\Delta{B}<0.2$, 2694 with $\Delta{B}<0.1$); 250 (183, 111)
cases where three entries match; and 20 (13, 8) where 4 entries match.
The mean $B$-band photometric offset for these coincidences is
0.00$\pm$0.27\mg\ ($0.00\pm0.08$\mg, $0.00\pm0.05$\mg), and the mean
positional offset is $1.17\pm0.61''$ ($1.07\pm0.56''$, $1.03\pm0.55''$;
all quoted errors are standard deviations), suggesting that $\sim$5\%\ of
objects in the catalogue may be the result of spurious multiple entries.

\subsection{Adopted results}

\subsubsection{Astrometry}
\label{astrom_txt}

We conclude that the astrometry for all the datasets considered is
internally consistent to better than 0.5$''$ rms.  The offsets between
the APM astrometry and that from the photometric catalogues average
$\Delta\alpha\simeq{+0.4''}$, $\Delta\delta<0.1''$, APM minus
other.\footnote{The APM system currently available on-line has been
transformed to the Tycho-2 system} For reference, the APM sampling
interval -- `pixel size' -- is 0.5$''$, and the 2dF fibres project to
2$''$ diameter.  We adopt the APM astrometry for the main 2dF catalogue.

\subsubsection{Photo{\-}metry}
\label{photom_txt}

A comparison of photometric results for the stars in the spectroscopic
survey is shown in Fig.~\ref{Phot1}.  The distribution of APM
photometric residuals is highly asymmetric for all reference
catalogues, with an extended tail where the APM magnitudes are
brighter than the comparison catalogue values. This asymmetry renders
the arithmetic mean offset a statistic of limited utility; even the
median is skewed by the moderately large fraction of these large
residuals.  The mode therefore gives the most meaningful average
measure of the offsets for the APM photo{\-}metry.

We made a small ($\mathcal{O}[0.1\mg]$) zero-point adjustment to the
original \BJ\ photo{\-}metry to bring the modal offset from MCPS $B$ to
zero (colour terms being negligible).  After this correction, the
modal offset for OGLE is $-0.03\mg$ (APM fainter), and for \citeauthor{pm02}
$-0.13\mg$.  For matches within the 2dF sample, we also find median
offsets MCPS$-$OGLE = +0.02\mg\ (1265 stars, excluding those for which
\citeauthor{pm02} photo{\-}metry is substituted in the MCPS), MCPS$-$\citeauthor{pm02} =
+0.13\mg\ (2396), and OGLE$-$\citeauthor{pm02} = +0.15\mg\ (1137).  On the basis
of these comparisons, we conclude that the APM $B$ magnitudes have a
zero-point uncertainty of order $0.1$\mg.  Characterizing the (large)
statistical uncertainty in the APM photo{\-}metry in a useful way is
complicated by the asymmetric nature of the distribution of magnitude
differences. As an indicator, we find that 68\%\ of matches have
$B$-magnitude offsets in the range $\sim$+0.6\mg/$-$0.4\mg\ of the mode.

\begin{figure} 
\begin{center}
\includegraphics[width=230pt]{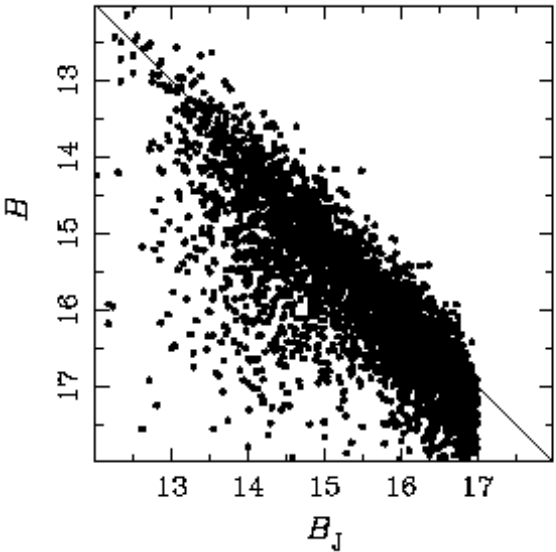}
\caption{Comparison of input APM \BJ\ photo{\-}metry and adopted $B$
photo{\-}metry.}
\label{Phot2}
\end{center}
\end{figure}

The OGLE and MCPS results are in good mutual agreement, with \citeauthor{pm02}'s
results $\sim$0.1\mg\ brighter than OGLE/MCPS for this sample.  (All
three datasets are tied to \citet{lan92} standards.)
\citet{pm02} notes essentially the same offset in a comparison with traditional
photoelectric photo{\-}metry.  For our present purposes, very precise
photo{\-}metry is not required, but for specificity (e.g., for diagrams),
we generally adopted results from the multi-epoch, high-redundancy
MCPS or OGLE surveys for preference (taking the better match to \BJ\
where necessary, to minimise mismatches).  The adopted main-catalogue
photo{\-}metry has 2513 values from MCPS, 747 from OGLE, 646 from
\citeauthor{pm02}, and 147 from APM.  The UIT-target photo{\-}metry comes
from MCPS and OGLE (97 and 10 stars, respectively).  The complete list
of cross-matches with photometric catalogues is given in
Table~\ref{catalogue} (Appendix~\ref{app_cat}), available in full
on-line, and a comparison of APM
\BJ\ magnitudes with adopted $B$ magnitudes is made in Fig.~\ref{Phot2}.

\section{The relationship between \Hg\ and absolute magnitude}
\label{MBWL}

\subsection{Overview}
\label{mbwlover}

\begin{figure*} 
\begin{center}
\includegraphics[width=145mm]{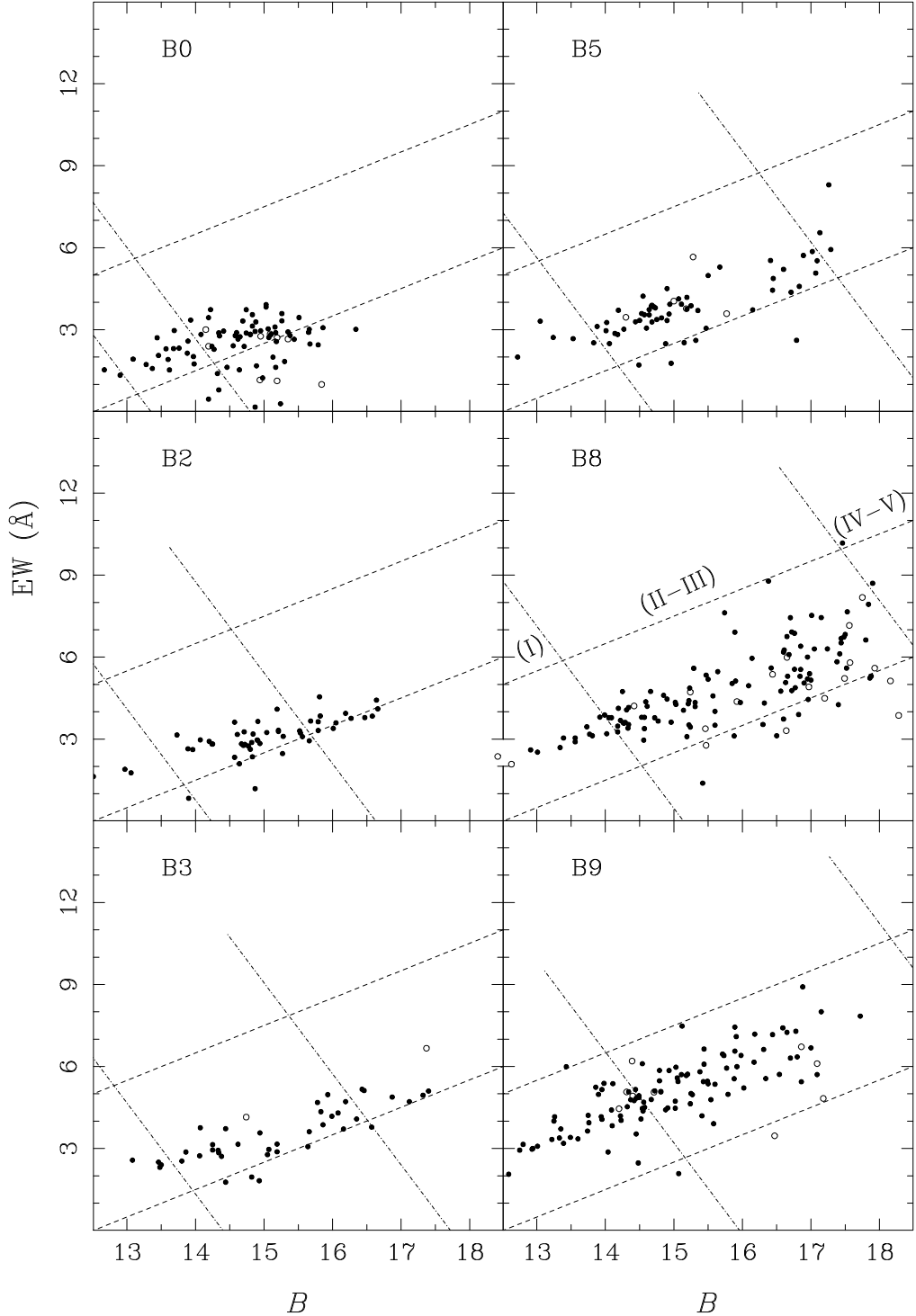}
\caption{\Hg\ equivalent width as a function of $B$ magnitude for 
B stars in the 2dF spectroscopic catalogue.  Only stars with `precise'
classifications (uncertainty $\pm$1 subtype) are shown.  The diagonal
lines of constant $(B-\Wl)$ (running lower left to upper right)
are intended solely to provide a point of
reference for the reader, and have no physical significance;  the
lines of constant $(B+0.3\Wl)$ show the adopted boundaries between
supergiants (luminosity class Ib and brighter), giants (II and III),
and main-sequence (IV and V) stars.
The open
circles show stars for which we have only APM photo{\-}metry, or for which
the CCD results differ by more than 1\mg\ from APM measurements (cf.\
Section~\ref{pherr_txt}).  The brightest B8 star shown (just outside
the main data frame) is AzV~72, which has previously been noted as a
member of the class of SMC supergiants supposedly having anomalously
strong hydrogen lines.  It appears unexceptional in our dataset (cf.\
Section~\ref{anom_txt}).
\label{EWB_B}}
\end{center}
\end{figure*}

\begin{figure*} 
\begin{center}
\includegraphics[width=145mm]{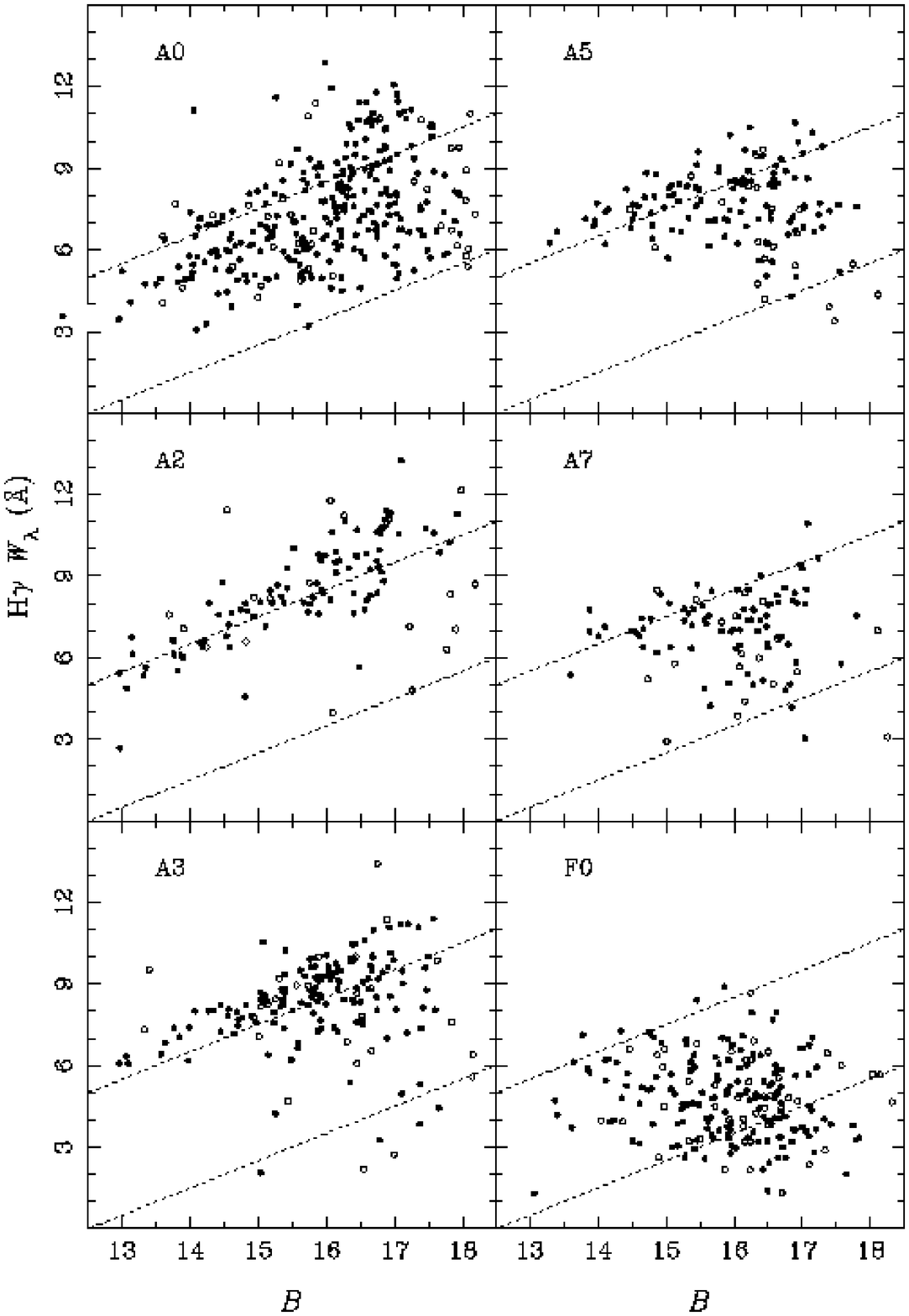}
\caption{\Hg\ equivalent width as a function of $B$ magnitude for 
A stars in the 2dF spectroscopic catalogue.  
Only stars with `precise'
classifications (uncertainty $\pm$1 subtype) are shown.
The diagonal dotted lines of constant $(B-\Wl)$
are intended solely to provide a point of reference for the reader,
and have no physical significance.
The open circles show stars for
which we have only APM photo{\-}metry, or for which the CCD results differ
by more than 1\mg\ from APM measurements (cf.\ Section~\ref{pherr_txt}).
The brightest A0 star shown (just outside the
main data frame) is Sk~1, which has previously been noted as a
member of the class of SMC supergiants supposedly having anomalously
strong hydrogen lines.  It appears unexceptional in our dataset (cf.\
Section~\ref{anom_txt}).
\label{EWB_A}}
\end{center}
\end{figure*}

Figures~\ref{EWB_B} and~\ref{EWB_A} show the relationship between \Hg\
equivalent widths, \Wl, and adopted $B$ magnitudes, as a function of spectral
subtype.  As expected, there is a clear general trend of increasing
\Wl\ with decreasing brightness, with $\partial{\Wl}/\partial{B}
\simeq +1$\AA/\mg\ (for all BA subtypes, $B \simeq 13.5$--17.5).
At $B \sim 14.5$, line strength increases slowly through the B
spectral sequence, from $\sim2.5$\AA\ at B0--B3 to $\sim4.5$\AA\ at
B9; $\Wl\simeq 7$--8\AA\ for all A subtypes, excepting the
transitional subtype A0, where $\Wl\simeq 6$\AA\ (with considerable
scatter), and A7, where there is a suggestion that \Wl\ starts to fall
off.  These results are in general agreement with previous work
\citep{hut66, bc74, azzo87}, though based on better-sampled,
higher-quality material.  The exception to this generalization is that
\Wl\ begins to decline at A7/F0 in our data, whereas \citet{azzo87}
finds monotonically increasing \Wl\ from early~B through early~F
subtypes, with $\Wl \simeq 8$\AA\ at F0$\;$Ib (at $B \simeq 15$).

The insensitivity of \Wl\ to A~spectral
subtype, throughout the magnitude range we sample, provides
justification for the subtype-independent luminosity classification
scheme adopted in Section~\ref{gal_cal} (although \Hg\ equivalent
widths are more strongly subtype dependent at fainter luminosity
classes).

\subsection{The dispersion}

There is significant scatter in the equivalent widths at a given
subtype and magnitude.  Observational errors could contribute to this
scatter in several ways, including photometric errors,
equivalent-width errors, and classification errors.

\subsubsection{Photometric errors?}
\label{pherr_txt}

Although the formal photometric uncertainties in the CCD data are
generally negligible compared with the observed dispersion,
arbitrarily large errors can be introduced if the photometric
catalogue star is mismatched with the spectroscopic catalogue entry.
As we have shown, the APM photo{\-}metry is also subject to relatively
large uncertainties.  To investigate these issues we examined the
residual datasets after exclusion of data for which the CCD results
differ from the APM photo{\-}metry by 1\mg\ or more (thereby reducing the
probability of cross-catalogue mismatches), or for which we
have only APM photo{\-}metry.  Excluding these arguably questionable data
reduces the scatter slightly, but the dispersion is large even for the
remaining CCD results (Figs.~\ref{EWB_B} and~\ref{EWB_A}).

\begin{figure*}
\begin{center}
\includegraphics[height=150mm, angle=270]{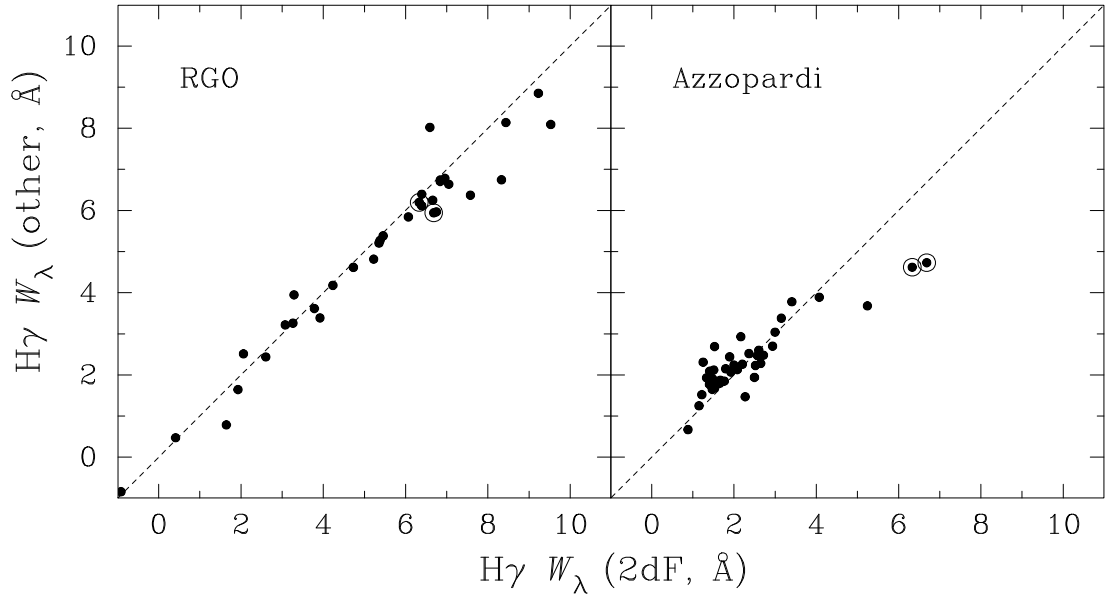}
\caption[]{Comparison of \Hg\ equivalent widths measured in 2dF and RGO
spectra, and by \citet{azzo87}.  
The circled points in each
panel show measurements for the same two stars, AzV~401 and~431;
the agreement between 2dF and RGO measurements suggest that it is
Azzopardi's measurements that are discrepant for these two stars.}
\label{HGcomp}
\end{center}
\end{figure*}

\subsubsection {Equivalent-width errors?}  
\label{ewerq}

Because the sky backgrounds in the 2dF spectra are not determined
directly, but are scaled from separate sky fibres, the accuracy of
background subtraction (and hence equivalent widths) is a source of
concern.  In practice, the continuum sky background (typically,
equivalent to a $\sim$20\mg\ star at $B$) is unlikely to exceed
$\sim$10\%\ of the gross signal, even in relatively poor cases.
Nonetheless, to check this point we made measurements on the
conventional long-slit spectra described in Section~\ref{longslit}.
The comparison of equivalent widths is made in Fig.~\ref{HGcomp};
agreement is good (although the comparison is limited to relatively
bright stars).

A small number of the brightest stars in our 2dF survey are in common
with the \citet{azzo87} survey of \Hg\ line strengths in SMC
supergiants.  We compare our equivalent widths with his in
Fig.~\ref{HGcomp}.  Agreement is generally acceptable, excepting three
outliers, for which our equivalent widths are noticeably greater than
his.  Two of those three outliers were included in the long-slit
observations (cf.~Fig.~\ref{HGcomp}), and our measurements suggest
that it is the newer data that are the more reliable.  (Azzopardi's
measurements were made on photographic objective-prism spectra
obtained with a resolving power perhaps half that of the 2dF
observations.)

Although the evidence therefore suggests that our measurements are
generally satisfactory, the tendency for some equivalent widths to
appear `too small' (Figs.~\ref{EWB_B} and~\ref{EWB_A}) does leave open
the possibilities that \Hg\ may, in those cases, be contaminated by
nebular emission at a level too small to be obvious at our resolution,
or that an overcorrection for the background signal has been applied.
The former issue is partially addressed by the identification in
Table~\ref{catalogue} of stars associated with catalogued
nebulosities, and by scrutiny of the Balmer-line spectra
(Section~\ref{Bem_stat}); and the latter by the preceding discussion.

\subsubsection {Classification errors?}   

Since the mean equivalent width of \Hg\  varies
as a function of spectral subtype at given $B$ magnitude, errors in
subtype assignments could introduce scatter.  The scatter is
noticeably large at spectral type A0, where the rate of change of \Wl\
with spectral type is greatest.  Late-B stars have significantly
smaller equivalent widths, on average, than the early-A stars, so
late-B stars misclassified as early-A stars could explain the tail of
`low-\Wl' A0 stars.

\begin{figure*} 
\begin{center}
\includegraphics[height=150mm,angle=270]{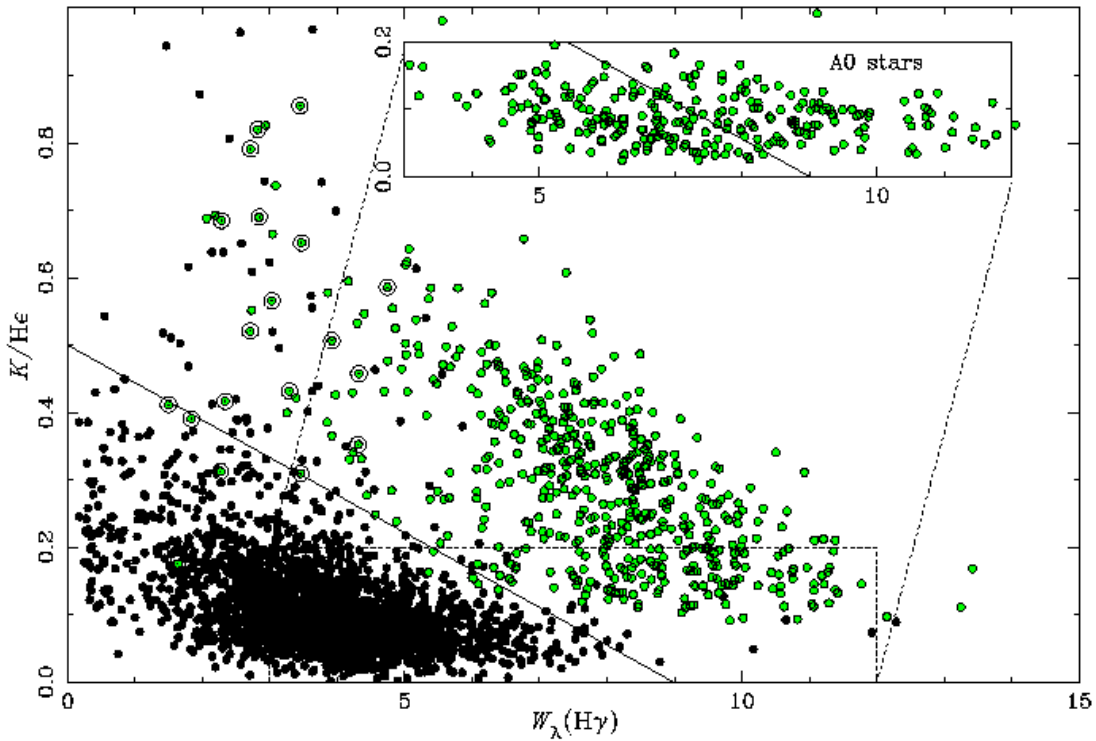}
\caption{
Ratio of Ca$\;K$ to H$\epsilon$+Ca$\;H$ as a function of \Hg\
equivalent width (in \AA).  The diagonal line is drawn `by hand' to
separate stars classified as B (solid black circles, lower left) from
those classified as A2--A7; these two groups are well separated in
this plane.  The displaced inset shows the A0 stars.  The circled
points are the possible `composite-spectrum' stars (Section~\ref{AFcomp}).
\label{HKvHg}}
\end{center}
\end{figure*}

To investigate this further, and more generally, we measured the
ratio of Ca$\;K$ to H$\epsilon$+Ca$\;H$ using a simple constrained
gaussian fit.  The equivalent-width ratio\footnote{This quantity is
closely related to, but not the same as, the $K$/H ratio used in the
visual classifications.  There is reasonable, though not perfect,
separation of individual A subtypes in this plane.}  is plotted
against \Hg\ equivalent width in Fig.~\ref{HKvHg}.  There is very good
separation between A- and B-type stars, {\em except} for
the A0 stars, which in this plane are mingled with both earlier and
later types.  In effect, for both B-type and A0~stars the $K$/H ratio
is just `small', while the \Hg\ equivalent width is influenced by
luminosity as much as by temperature.

The primary criterion for resolving the B9/A0 ambiguity is the
presence of \HeI~$\lambda$4471 in the earlier type.  A plausible
explanation of the scatter in \Wl, at A0 in particular, is, therefore,
that B9 spectra have been misclassified in poor-quality spectra with
S/N too low to allow the helium line (weak at this subtype) to be
identified.  An attempt to improve A0--B9 separation using automated
measurements of 4471/4481 proved fruitless, because of the low S/N of
the data.  Outliers in the \Wl--$B$ plane were therefore identified
and the spectra re-inspected for evidence of any anomalies.  For the
earlier spectral types, an anomalous position in this plane correlates
with Balmer-line emission (stellar or nebular); and at later types,
stars with emission are also displaced.  However, most mid- to late-A
outliers appear unexceptional (other than in respect of their
Balmer-line weakness) at our dispersion and signal-to-noise.

A further possibility is that unresolved companions may cause or
exacerbate the spread in the \Wl--$B$ plane.  This must be a factor at some
level, but comparison of the H$\gamma$ equivalent widths (at a given
spectral type) for cluster members and field stars reveals no greater
dispersion for the former, such as might be expected if blends were
important.  Furthermore, any close binary companions to BA giants and
supergiants are not expected to be significant contaminators of the
optical spectra, in general.  We conclude that the dispersion in
\Wl\ most probably arises primarily from a combination of measurement
errors, unresolved nebular emission, and minor classification errors,
with astrophysical effects (such as multiplicity and intrinsic emission) as
secondary contributors.

\begin{table*}
\begin{center}
\caption[]
{Numbers of stars by spectral type from the APM- and UIT-selected samples.  In the latter case,
totals are given for both the
pure UIT-selected sample and (in square brackets) other UIT targets which were matched in the
2dF catalogue.  `Em' lists the percentage of stars that show
emission at \Ha\ in each spectral bin (Section~\ref{Bem_stat}). `AF' is used for
the possible `composite-spectrum' stars (Section~\ref{AFcomp}).}
\label{sp_summary}
\begin{tabular}{crr c rrrrrrrc}
\hline
Spectral&\multicolumn{2}{c}{APM}&$\quad$&\multicolumn{4}{c}{UIT}&$\quad$&\multicolumn{1}{c}{Total}&$\quad$&Em.\\
Type    &\multicolumn{1}{c}{$N$}&\%&&\multicolumn{1}{c}{$N$}&\%  &\multicolumn{1}{c}{$N$}&\% &&\multicolumn{1}{c}{$N$}&& \%   \\
O+W            &   121+1\pA  &   {\pA}3  &    &    18  &   17  &   [21+1]  &    6  &    &    139+1 &&  20\\
B$<$5          &   2454\ppB  &       61  &    &    72  &   67  &   [274]   &   79  &    &     2526 &&  26\\
B$\ge$5        &    327\ppB  &   {\pA}8  &    &     9  &    8  &   [30]    &    9  &    &      336 &&  14\\
A$<$5          &    593\ppB  &       15  &    &     7  &    7  &   [17]    &    5  &    &      600 &&  {\pA}2\\
A$\ge$5(+`AF') &     233+19  &   {\pA}6  &    &   0+1  &    1  &  [1+2]    &    0  &    &   223+20 &&  {\pA}4\\
FG             &    306\ppB  &   {\pA}8  &    &     0  &    0  &    [1]    &    1  &    &      306 &&  {\pA}0\\
Total          &  4054\ppB   &           &    &   107  &       &   347     &       &    &     4161 &&  \\
\hline
\end{tabular}
\end{center}
\end{table*}

\section{Objects of special interest}
\label{seren}

\subsection{Balmer-emission statistics}
\label{Bem_stat}
The blue-region spectra at our disposal are, for the most part,
inadequate to distinguish reliably between any intrinsic narrow Balmer
emission and nebular emission.  Nonetheless, in a number of cases
resolved or structured Balmer emission is identifiable.  As a rough
guide to objects likely to be of interest, we therefore categorized
all observed Balmer lines as showing absorption; narrow (i.e.,
unresolved, probably nebular) emission; resolved emission;
double-peaked emission; or P-Cygni profiles.

Stars were categorized rather conservatively (i.e., only reasonably
definite cases were categorized as `resolved' rather than `narrow', or
as `double' rather than `resolved').  Each Balmer line was categorized
independently for each star (4161 targets have \Hg\ observed, 2095
\Hb, and 1091 \Ha).  After reviewing a  handful of anomalies
(and consequently revising \Hg\ `r?' to `n'), all stars having red
spectra and showing intrinsic
\Hb\ emission (or suspected intrinsic emission) also show
\Ha\ emission, and all stars showing intrinsic \Hg\ emission show
\Hb\ and \Ha\ emission, where data exist.

A hundred and ninety-seven stars with usable red-region 2dF spectra
(18$\%$) show intrinsic \Ha\ emission of some sort; 161 stars
with \Hb\ spectra (8$\%$) show \Hb\ emission; and 154 stars (4$\%$) show
\Hg\ emission.  The targets for the red-region spectra
were biassed towards objects with `interesting' blue spectra;
correcting for this bias, $\sim$14$\%$\ of stars in the input
catalogue show \Ha\ emission.  To a good approximation, in our data
\Ha\ is twice as effective as \Hb\ in disclosing intrinsic Balmer
emission, and \Hg\ half as effective.

In the majority of cases ($\sim$80$\%$), blue-region Balmer emission
consists of resolved (or marginally resolved), undisplaced (or only
moderately displaced) narrow emission, without obvious emission in any
other lines.  Emission features in the \Ha\ spectra are invariably
substantially broader, or clearly structured.  Table~\ref{sp_summary}
includes a summary of the distribution of Balmer emission by spectral
type, peaking at early~B.

\subsection{Be-type spectra}
\label{Bem_txt}

\begin{figure*}
\begin{center}
\includegraphics[width=150mm]{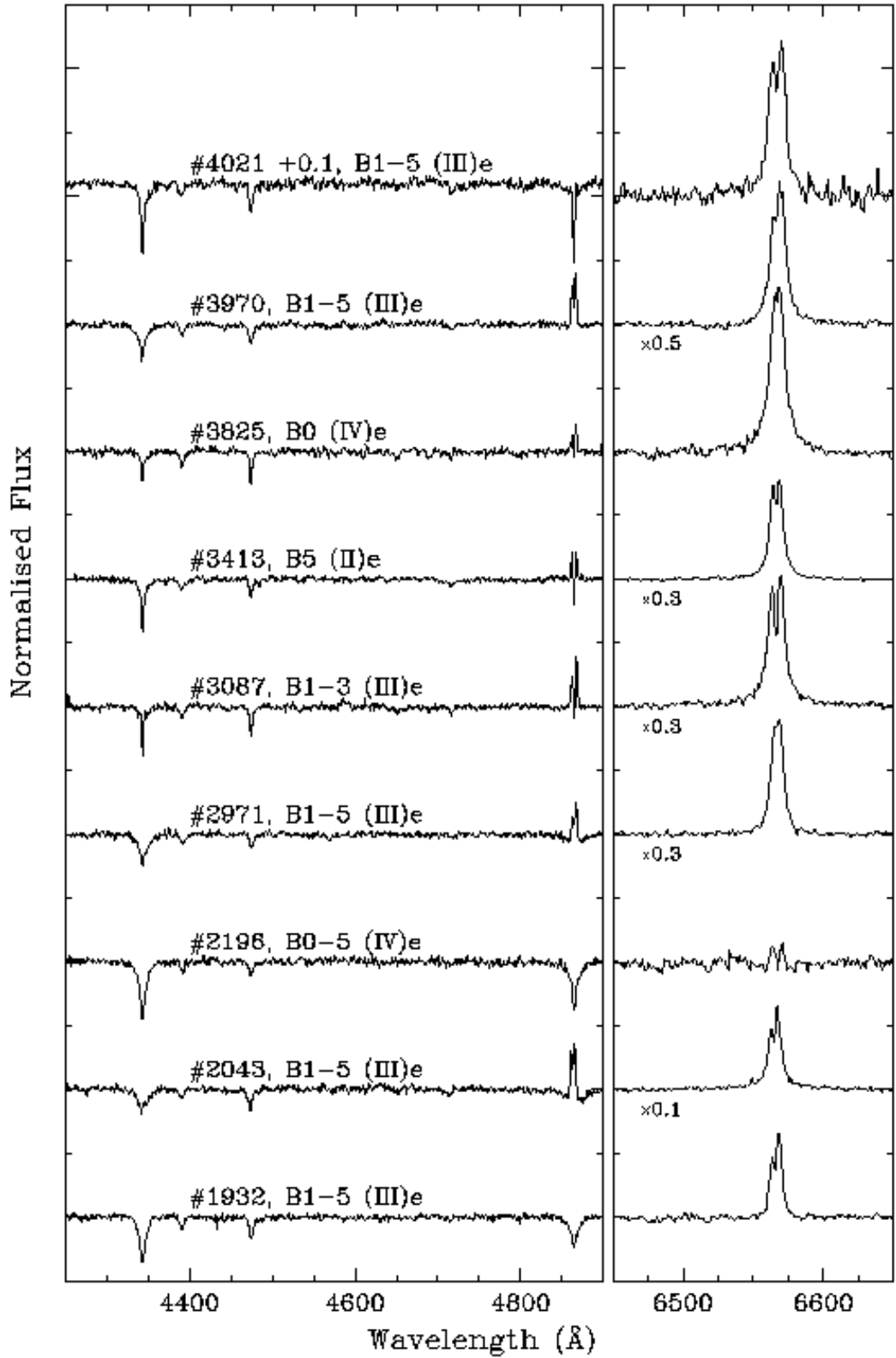}
\caption[] {Spectra of selected Be stars.  
Each spectrum is vertically offset by 1.0 continuum units.  The
luminosity classes assigned to these stars should be regarded with
caution, as Balmer-line emission is liable to result in a `too bright'
classification with our `hybrid' methodology (Section~\ref{blum}).}
\label{be_stars}
\end{center}
\end{figure*}

BA stars reliably exhibiting double-peaked emission in at least one
Balmer line (or suspected double-peaked \Hb\ accompanied by \Ha\
emission) are listed in Table~\ref{Bem_tab}, and illustrative spectra
are shown in Fig.~\ref{be_stars}.  The stars with double-peaked
emission show a spread in absolute magnitude from $M(B) \simeq -5$ to
the cutoff of our survey ($M(B)
\simeq -1$), with a mode of about $M(B) \simeq -3$, and are
classified (usually rather coarsely) as early B, and so are likely to
represent the bright tail of the distribution of classical Be stars,
if we adopt the {\em Hipparcos}-based absolute-magnitude scale for
Galactic Be stars given by \citet{weg00}.  Because of their relative
brightness, these 2dF stars are attractive targets for, e.g.,
investigation of short-term line-profile variability
\citep[{cf.}][]{baa02}.

Encouraged by our referee, we have also listed stars whose spectra
display single-peaked, resolved emission at either H$\alpha$ or
H$\beta$ as `Be'-type in Table~\ref{catalogue}.  Many or most of these
stars may be classical Be stars, but our primary motivation in adding
the `e' suffix to the catalogue spectral types is to alert users to the
qualifying supplementary emission-line information.

None of the stars in Table~\ref{Bem_tab} have been explicitly
identified as Be stars in the literature, but several other previously
known objects are included in the 2dF catalogue: one from the sample
of Be stars studied by \citet{humm99}, 2dFS\#1115 (NGC~330
KWB$\;$Be~258, Grebel~9), and a further four potential Be stars listed
by \citet*{kwb99}: 2dFS\#1077=KWB$\;$Be~278; \#1087=522; \#1277=122;
and \#1326=355.  Of these five stars, 2dFS\#1077 and \#1277 show
marginally resolved (i.e., broadened) emission at \Hg\ (only
2dFS\#1326 has \Hb\ observed), while 2dFS\#1277 and~\#1326 have red
spectra, both showing strong, single-peaked \Ha\ emission.  

\begin{table*}
\begin{center}
\caption[]
{Presumed classical Be stars: B stars showing double-peaked emission
in the Balmer lines.  The last three columns encode the spectral
appearance in \Hg, \Hb, and \Ha.  Here `a' means absorption, `n' means
narrow (possibly nebular) emission, `r' means resolved (though narrow)
emission, `d' means double-peaked emission, and `e' means broad
single-peaked emission.  The luminosity classes assigned to these
stars should be regarded with caution, as Balmer-line emission is
likely to result in a `too bright' classification with our `hybrid'
methodology (Section~\ref{blum}).}
\label{Bem_tab}
\begin{tabular}{clclllll}
\hline
2dFS& 
$\quad$Other&
$B$ &
$\quad${Sp.}&
\multicolumn{3}{c}{Balmer}\\
\hline
0080     &                   & 17.18  & B0--5 (V)e         &   $\gamma$a & $\beta$d    &                         \\
0113     &                   & 15.64  & B1--2 (III)e       &   $\gamma$a  & $\beta$d    &                         \\
0306     &                   & 14.35  & B0--5 (II)e        &   $\gamma$a? & $\beta$d    &                         \\
0922     &                   & 16.00  & B0--5 (V)e         &   $\gamma$d? & $\beta$d    &                                  \\
1306     &                   & 16.62  & B0--5 (IV)e        &      $\gamma$a    &&  $\alpha$d                   \\
1496     & MA93-1216         & 15.77  & B0.5 (V)e           &      $\gamma$d?   & & $\alpha$e                   \\
1551     &                   & 16.51  & B0--5 (IV)e       &      $\gamma$a    & & $\alpha$d                   \\
1558     &                   & 17.44  & B0--5 (V)e          &   $\gamma$a & $\beta$d    &                         \\
1932     &MA93-1490          & 15.32  & B1--5 (III)e       &   $\gamma$a & $\beta$a    &  $\alpha$d                   \\
2043     &MA93-1539          & 15.99  & B1--5 (III)e       &   $\gamma$a & $\beta$d    &  $\alpha$d                   \\
2196     &                   & 16.21  & B0--5 (IV)e        &   $\gamma$a & $\beta$a    &  $\alpha$d                   \\
2267     &MA93-1621          & 15.84  & B1--5 (III)e       &      $\gamma$d?   &&  $\alpha$d                   \\
2316     &AzV~422, MA93-1637 & 13.97  & B1--5 (II)e            &   $\gamma$a & $\beta$d    &                         \\
2360     &                   & 16.96  & B0--5 (V)e          &   $\gamma$a & $\beta$d    &  \\
2598     & MA93-1707         & 16.25  & B1--5 (III)e        &      $\gamma$a   & &  $\alpha$d                   \\
2613     & MA93-1711         & 15.65  & B0--5 (III)e        &   $\gamma$a & $\beta$d?   &  $\alpha$d             \\
2786     &                   & 17.53  & B1--5 (V)e         &      $\gamma$a    &&  $\alpha$d                   \\
2794     &                   & 15.02  & B1--5 (II)e            &   $\gamma$a & $\beta$a    &  $\alpha$d   \\
2802     &                   & 15.66  & B0--5 (III)e       &   $\gamma$n & $\beta$r    &  $\alpha$d                   \\
2971     & MA93-1775          & 15.83  & B1--5 (III)e       &   $\gamma$a & $\beta$d    &  $\alpha$e                   \\
2986     &                    & 15.07  & B1--3 (III)e          &   $\gamma$a & $\beta$a    &           $\alpha$d          \\
3087     & MA93-1813          & 15.78  & B1--3 (III)e      &   $\gamma$d & $\beta$d    &  $\alpha$d                   \\
3325     &                    & 16.07  & B0--5 (IV)e            &      $\gamma$a    &&  $\alpha$d                   \\
3413     & MA93-1868          & 14.96  & B5 (II)e           &   $\gamma$a & $\beta$d    &  $\alpha$d                   \\
3426     & MA93-1871          & 16.15  & B0--5 (III)e        &   $\gamma$a & $\beta$d    &                       \\
3436     &                    & 15.93  & B1--5 (III)e        &   $\gamma$a & $\beta$d?   &  $\alpha$d                   \\
3479     &  MA93-1881         & 15.82  & B1--5 (III)e        &      $\gamma$a    &&  $\alpha$d                   \\
3512     &  MA93-1886         & 14.82  & B3 (II)e              &      $\gamma$a    &&  $\alpha$d                   \\
3573     &  MA93-1894         & 15.88  & B1--3 (III)e        &      $\gamma$r    &&  $\alpha$d                   \\
3628     &                    & 14.67  & B1--3 (II)e         &      $\gamma$d?   &&  $\alpha$e                   \\
3716     &                    & 14.97  & B1--5 (II)e            &      $\gamma$a    &&  $\alpha$d                   \\
3730     &                    & 16.99  & B0--5 (V)e          &   $\gamma$d? & $\beta$d    &                         \\
3795     &                    & 15.15  & B1--3 (III)e        &      $\gamma$r?   &&  $\alpha$d                   \\
3825     &                    & 15.19  & B0 (IV)e           &   $\gamma$a & $\beta$d    &  $\alpha$e                   \\
3928     &                    & 15.36  & B1--5 (II)e         &   $\gamma$d? & $\beta$d    &                         \\
3970     &                    & 15.48  & B1--5 (III)e       &   $\gamma$a & $\beta$d    &           $\alpha$d          \\
3998     &                    & 15.49  & B0--5 (III)e           &      $\gamma$a    &&  $\alpha$d                   \\
4021     &                    & 15.49  & B1--5 (III)e           &   $\gamma$a & $\beta$d?   &  $\alpha$d                   \\
\hline
\end{tabular}
\end{center}
\end{table*}

\subsection{B[e] stars}
\label{Ae_stars}

2dFS\#2837 has previously been catalogued as LHA~115-N82, Lin~495, and
MA93-1750 (\citealt{hen56}; \citealt{l61}; \citealt{ma93}), and was
reported as a B[e] star by \citet{hey90}.  It is the only previously
reported B[e] star in our sample.  We observed it in 1998 with 2dF, and
again in 2001 with the RGO spectrograph.  The two spectra are
indistinguishable (the RGO spectrum is shown in Fig.~\ref{odd_fig}),
and similar to that shown by \citeauthor{hey90} if allowance is made for
the differences in resolution.

The only other star in the 2dF dataset that shows similiarities to
\#2837, and in particular the same Fe$\;${\sc [ii]} emission lines, 
is 2dFS\#1804.
Unfortunately, we have only a single, blue-region, observation of this
star (Fig.~\ref{odd_fig}), but, like \#2837, it has been repeatedly
catalogued as an \Ha\ emission object: LHA115-S38, Lin~418, MA93-1405
(\citealt{hen56};
\citealt{l61}; \citealt{ma93}).  This relatively bright star ($B
\simeq 14.0$) is clearly worthy of further study.

\subsection{Composite-spectrum targets}
\label{AFcomp}

\begin{figure} 
\begin{center}
\includegraphics[width=230pt]{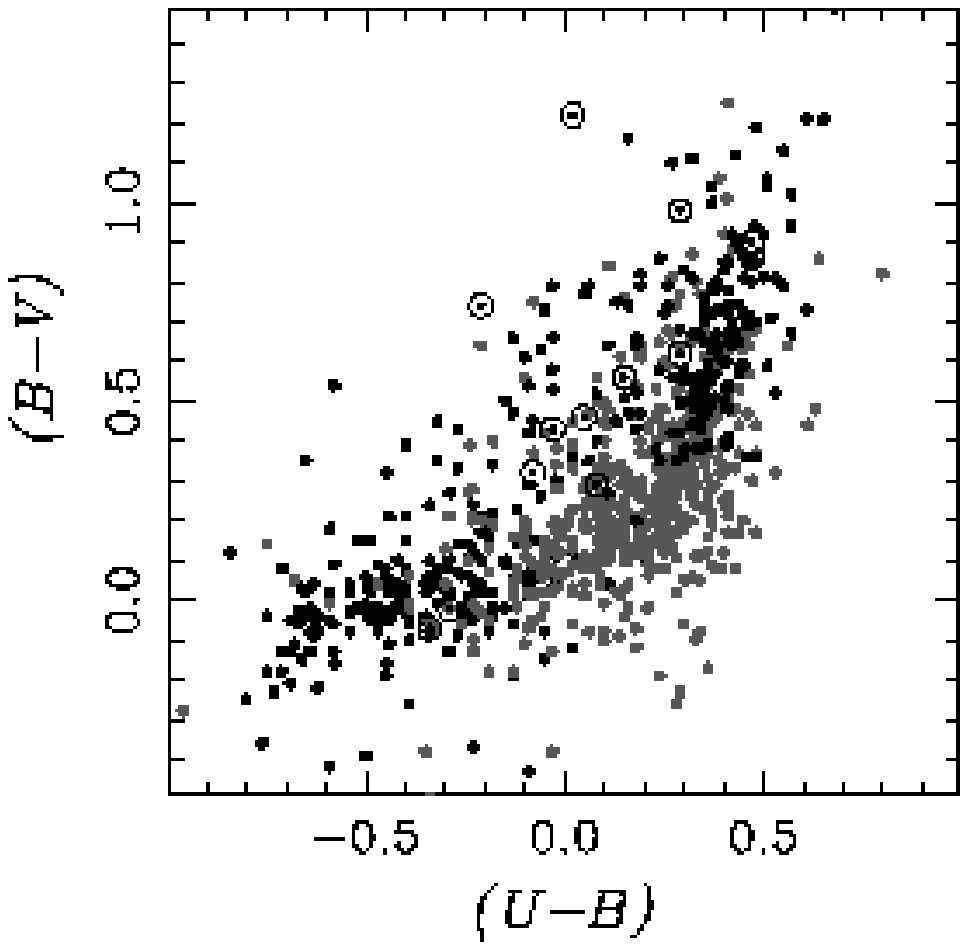}
\caption{The two-colour diagram for late-B, A, and F stars, based on
photo{\-}metry from \citet{pm02}.
The possible `composite-spectrum' stars (Section~\ref{AFcomp}) are shown as
circled dots, and apparently normal A stars in a different shade to the
BF stars.}
\label{Am2C}
\end{center}
\end{figure} 

Our examination of eighteen of the 2dF spectra indicated an
$\sim$A-type classification from Ca$\;K$, while other metal lines
(including the $G$ band) suggested spectral type~$\sim$F.  These
targets form a distinct group in the SMC survey; the spectra are
superficially suggestive of Galactic Am~stars in several cases, but
classical Am stars would be too faint to feature in our survey.  Our
targets are unlikely to be foreground objects, both because of the
observed radial velocities and because the surface density is
implausibly high.  We also provisionally rule out contamination by the
solar (i.e., twilight or bright-moon) spectrum; most of these stars
were observed in dark skies when the moon was below the horizon (and
similar features are not seen in any B-star targets).

We conclude that many of these observations may represent composite
spectra.  The rate of incidence is $\mathcal{O}(1\%)$, similar to
Galactic values, but these cannot be exact counterparts of Galactic
composite-spectrum systems, which typically consist of a
near-main-sequence A~star with a G~giant; our targets are somewhat
more luminous.  They are, however, `too blue' in $(U-B)$ and/or `too
red' in $(B-V)$ for the most part (Fig.~\ref{Am2C}), consistent with a
generic composite-spectrum intepretation.  Unfortunately, the
photometric anomalies are not sufficiently clear-cut for these stars
to be uniquely or unambiguously identifiable from colours alone.  In
the H/$K$ plane (Fig.~\ref{HKvHg}) these stars are also outliers,
falling at the small \Wl(\Hg) end of the A-star distribution (which
could, however, be construed as a consequence of incorrect background
subtraction).

R.O. Gray has extensive experience with spectral classification in the
relevant part of the HRD (e.g., \citealt{gnw01}), and very kindly
volunteered to examine some of these spectra.  He was able to
re-interpret three of the targets as having spectra similar to normal
or near-normal Galactic F-type stars, within bounds of the data quality
(although the radial velocities argue against this interpretation),
but confirmed apparent abnormalities in several other cases.  We have
no repeat spectroscopy for any of these stars, and so are unable to
rule out some instrumental, rather than astrophysical, peculiarity;
attempts to obtain new spectra were thwarted by bad weather.
Good-quality, long-slit spectra of just one or two of these targets
could go a long way towards clarifying their true nature.

\begin{figure*}
\begin{center}
\includegraphics[height=180mm, width=160mm, angle=0]{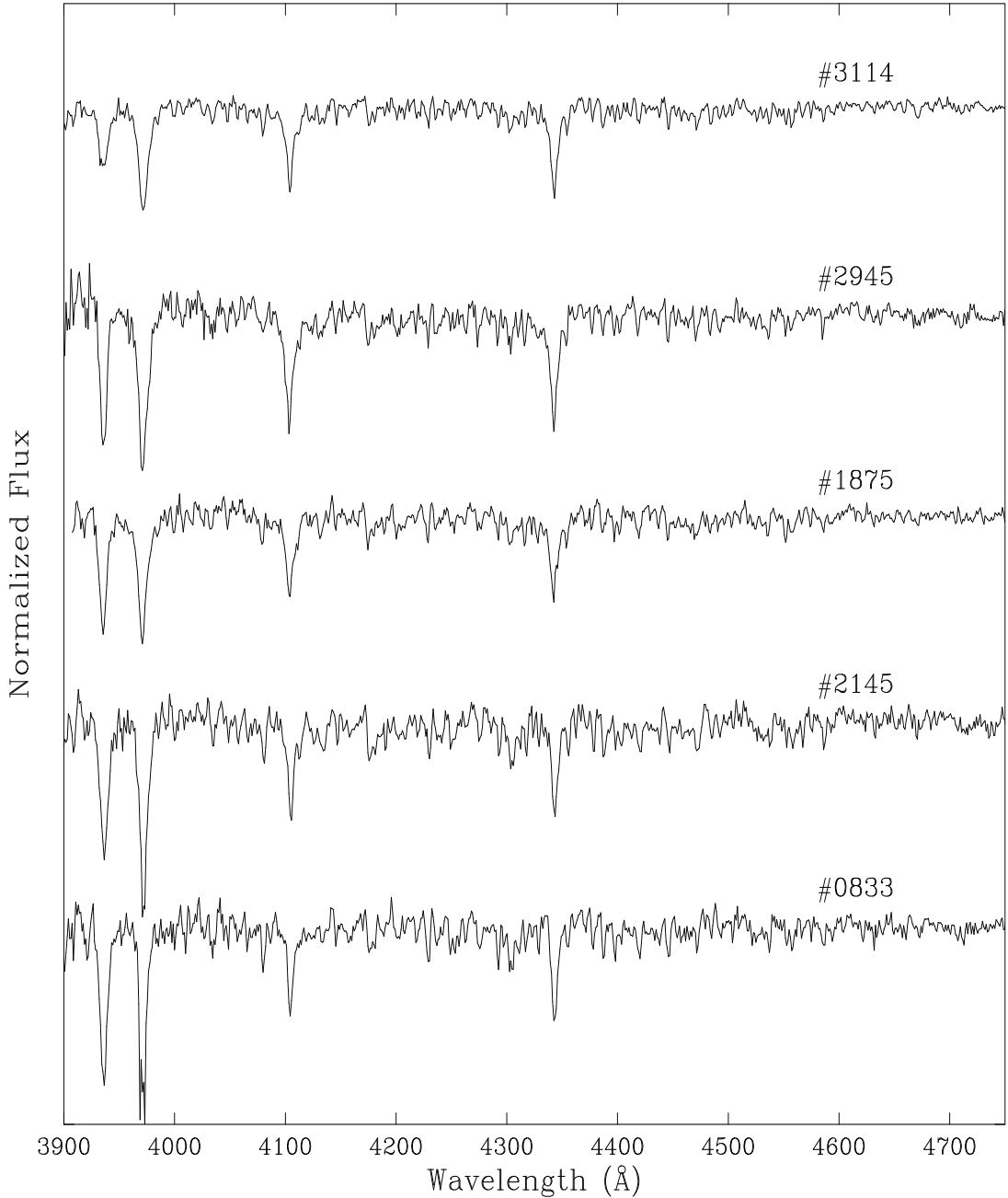}
\caption[]
{2dF SMC spectra -- VI: Examples of possible composite-spectrum
targets.  Stars are identified by 2dFS catalogue number.  Successive
spectra are vertically offset by one continuum unit.  }
\label{Am_fig}
\end{center}
\end{figure*}

\subsection{The `Anomalous A supergiants'}
\label{anom_txt}

Following earlier work on LMC stars by \citet{sa72} and \citet{fd72},
\citet{azzo82} and \citet{h83} noted a population of `anomalous
A-type supergiants' in the SMC, characterized by unusually large
Balmer-line strengths and red $(U-B)$ colours.  In our sample (for
which the great majority of targets are substantially fainter and less
luminous that those studied by \citeauthor{azzo82} and by
\citeauthor{h83}) there is
very little evidence for any stars having these characteristics
(cf.\ Figs.~\ref{EWB_B}, \ref{EWB_A}, and~\ref{Am2C}).

We have only two targets among those listed by \citet[][\,her
Table~1C]{h83} as anomalous: Sk~1 (2dFS\#0109) and AzV~72 (Sk~37,
2dFS\#0765).  Our classifications (A0$\;$(Ia) and B8$\;$(Iab),
respectively) are in excellent agreement with hers (A0$\;$I and
B8$\;$I), but neither star is exceptional in the \Wl--$B$ plane in the
context of our dataset\footnote{For AzV~72 we have only APM
photo{\-}metry: $B=12.43$.
\citet{av75} and \citet{ard80} report concordant $UBV$ 
measurements giving a somewhat fainter $B \simeq 12.88$, strengthening
the normality of this star in the \Wl--$B$ plane.}  (Figs~\ref{EWB_B}
and~\ref{EWB_A}).  This is consistent with an unpublished study by
E.L.~Fitzpatrick (personal communication), in which he argues,
persuasively in our view, that many reported anomalies in Magellanic
Cloud A~supergiants can be attributed to the lack of a proper
reference framework of {\em normal} stars at Cloud metallicities (cf.\
\citealt{djl97} for a related critique of metal-line luminosity 
classification of SMC B-type supergiants).

\citet*{hkg91} proposed a physical explanation for the 
strength of the Balmer lines in the supposedly anomalous stars.  They
suggested that these stars are post-red-supergiants, with enhanced
surface helium abundances leading to increased atmospheric pressure at
given temperature.  By itself, this is a plausible argument, but
standard evolutionary models suggest that post-red-supergiant BA
supergiants should have around half the mass of pre-red-supergiants at
the same luminosity and \Teff.  The reduction in \logg\ would be
expected to lead to {\em reduced} Balmer-line strengths (which may be
a contributory factor to the small-\Wl\ stars in our sample).

\subsection{The UIT targets}
\label{uit_txt}

Classifications for the 107 UIT-selected targets are summarised in
Table~\ref{sp_summary}.  As for the main catalogue, roughly
three-quarters of the sample are B-type stars, but the UV-selected
sample is uncontaminated by FG stars, and features a much smaller
proportion of A-type stars than the APM, $(B-R)$-selected sample.
We also searched the main APM catalogue for UIT stars.  There are 347
matches with separations of less than 2$''$ (cp.\ 386 for 5$''$, 223
for 1.2$''$), and the distribution of spectral types closely matches
that of the UIT-selected sample (Table~\ref{sp_summary}), albeit with somewhat 
smaller fraction of O~stars.

\citet{parker02} discusses whether there was a hitherto undiscovered
population of early-type field stars in the Clouds, well away from
obvious OB associations.  The UIT-selected sample is clearly more efficient in
identifying candidate O-type stars than is using the APM data as an input
catalogue.  However, the UV-selected sample does not appear to reveal
a new and distinct O-star population; a significant majority of the
UIT targets observed with 2dF are early-B-type stars.

\subsection{Individual objects}
\label{odd_txt}

\begin{figure*}
\begin{center}
\includegraphics[width=150mm]{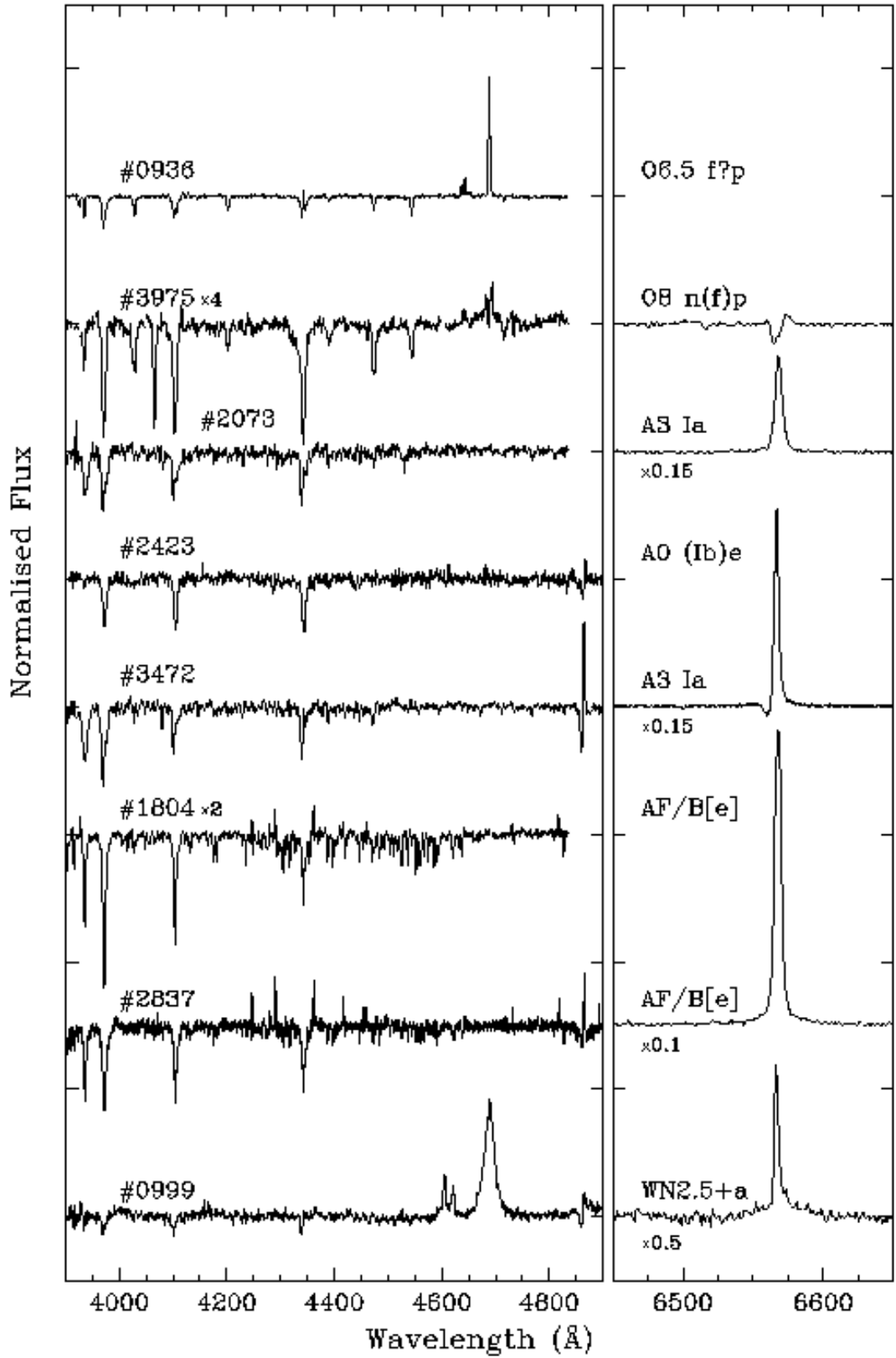}
\caption[ ] {Objects of particular interest, identified by 2dFS  
number; see Sections~\ref{Ae_stars}
and~\ref{odd_txt} 
for a discussion of the spectra shown.
Tickmarks on the $y$ axis are every
1.0 continuum units.}
\label{odd_fig}
\end{center}
\end{figure*}

\subsubsection{2dFS\#0936}
\label{odd0936}
The spectrum of star \#0936 has narrow,
strong He~\2 \lam4686 emission with an equivalent width of $-$3.5\AA\
($\pm$~0.2), and weaker N~\3 and C~\3 lines (see Fig.~\ref{odd_fig}).  
The spectrum is classified here as
O6.5$\;$f?p, a category first introduced by \citet{wal72}.  
The principal defining characteristic of Of?p stars is  C~\3
\lam4647-50-51 emission  comparable in intensity to  
N~\3 \lam4634-40-42, although the relative intensities of N~\3, C~\3
and He~\2 emission vary significantly in several well observed
examples \citep{yael, wh03}.  
Aside from the 
weaker C~\3 emission, the spectrum of \#0936 is similar to that of AzV~220
\citep{wal00};  in common with AzV~220, \#0936 is 
too faint (with $M(B) \simeq -5.0$) for a `normal' supergiant, when
compared to the calibration given by \citet{wal73}.  
The spectrum also displays Balmer emission (unresolved at \Hg), 
the origins of which are not clear; the star is in H35
\citep{h85}, which \citet{bs95} list as type ``AN -- an association
which shows some traces of H$\;${\sc ii} emission.''  

Further
observations of this star would be useful to distinguish between
intrinsic and nebular emission, and to monitor for spectral variations.
In the course of their survey for WR stars, \citet{pm01} observed
2dFS\#0936 (their ``Anon-1''), assigning a spectral type of
O5$\;$f?p. The 2dF spectrum (which has greater wavelength coverage) is
not consistent with this classification, since He~\1 \lam4026 is
stronger than He~\2 \lam4200.  At the current time it is unclear
whether this difference is attributable to differences in data quality
or to intrinsic spectral variability.

\subsubsection{2dFS\#0999: SMC-WR 9}

Star 2dFS\#0999 (Fig.~\ref{odd_fig}) was the ninth Wolf-Rayet (WR)
star discovered in the SMC
\citep*{mvd91}.  The strong N~\5 emission, in
combination with an absence of N~\4, gives the WN2.5 type
\citep{vdh96}, and the absorption
component is consistent with an O3 companion, based on the absence of
He~\1~\lam4471.  The spectrum has previously been classified as WN2.5$+$O5:
\citep{mvd91} and  WN3$+$O3--4 \citep{pm01}.
The 2dF spectrum has resolution comparable to Morgan's data, and
a comparison shows the N~\5 to
He~\2 \lam4686 ratio has apparently increased.

\subsubsection{2dFS\#3235: AzV~490/Sk~160}
\label{azv490}
The 2dF spectrum of 2dFS\#3235 (AzV~490; Sk~160 in the catalogue of
\citealt{sa68}) shows strong He~\2 emission at \lam4686.  This
is the optical counterpart of SMC~X-1.  The optical spectrum has
long been known to be variable \citep[e.g.,][]{oh74}, and our RGO
spectrum shows differences from the 2dF spectrum (e.g., He$\;${\sc
ii}~$\lambda$4686 double-peaked in the former, single-peaked in the
latter).  Our 2dF spectrum is very similar to that of HDE~269896 in
the LMC, classified as ON9.7$\;$Ia$+$ (\citealt{wal77},
\citealt{wf90}).  

\subsubsection{2dFS\#3472;  \#2073, \#2423}
\label{3472}

2dFS\#3472 (A3$\;$Ia) is an emission-line object, previously
catalogued as Lin~530 and MA93-1879 (\citealt{l61}; \citealt{ma93}).
RGO (H$\alpha$) and 2dF (blue-region) spectra are shown in
Fig.~\ref{odd_fig}.  Although Balmer-line emission is not particularly
rare in luminous Galactic A~supergiants, the P-Cygni H$\alpha$ line in
2dFS\#3472 is exceptionally strong, with an emission equivalent width
of $\sim$40\AA.

2dFS\#2423 (A0$\;$(Ib)) is the only other A-type star in our
catalogue showing clear P-Cygni emission at \Hb, albeit
rather weaker than in \#3472,
while 2dFS\#2073 (A3$\;$Ia) shows probable P-Cygni emission at \Hg,
together with strong \Ha\ P-Cygni emission  (Fig.~\ref{odd_fig}).

\subsubsection{2dFS\#3975: Sk~190}

The spectrum of star \#3975 (Fig.~\ref{odd_fig}) is classified here as
O8n(f)p.  The primary requirement for membership of this class is a composite
emission and absorption structure in the He~\2 \lam4686 line
\citep{wal73}.  The 2dF spectrum was both noisy and near the edge of
the engineering-grade chip in the 1998 observations, introducing
assorted cosmetic features; subsequent observation with the RGO
spectrograph confirmed the classification.  This is only the second Onfp
spectrum seen in the SMC, after AzV~80 \citep{wal00}.  Sk~190 has
previously been classified as O8 Iaf \citep{pm02}, suggesting the
possibility of time variability in the intensity and morphology of the
He~\2 \lam4686 line.

\section{Acknowledgements}
CJE was funded by PPARC during the course of this work.  This paper is
based on data obtained with the Anglo-Australian Telescope, and we
thank the staff of the observatory (particularly Russell Cannon) for
their support.  Ed Fitzpatrick, Richard Gray, and our referee, Nolan
Walborn, provided helpful comments, and we also thank Martin Cohen,
Danny Lennon, and Nidia Morrell for useful correspondence.

\bibliography{ME268rv}

\begin{thebibliography}{}

\bibitem[\protect\citeauthoryear{{Abt}}{{Abt}}{1981}]{abt81}
{Abt} H.~A.,  1981, \apjs, 45, 437

\bibitem[\protect\citeauthoryear{{Abt}}{{Abt}}{1985}]{abt85}
{Abt} H.~A.,  1985, \apjs, 59, 95

\bibitem[\protect\citeauthoryear{{Ardeberg}}{{Ardeberg}}{1980}]{ard80}
{Ardeberg} A.,  1980, \aas, 42, 1

\bibitem[\protect\citeauthoryear{{Azzopardi}}{{Azzopardi}}{1982}]{azzo82}
{Azzopardi} M.,  1982, in The Most Massive Stars p.~227

\bibitem[\protect\citeauthoryear{{Azzopardi}}{{Azzopardi}}{1987}]{azzo87}
{Azzopardi} M.,  1987, \aas, 69, 421

\bibitem[\protect\citeauthoryear{{Azzopardi} \& {Vigneau}}{{Azzopardi} \&
  {Vigneau}}{1975}]{av75}
{Azzopardi} M.,  {Vigneau} J.,  1975, \aas, 22, 285

\bibitem[\protect\citeauthoryear{{Azzopardi} \& {Vigneau}}{{Azzopardi} \&
  {Vigneau}}{1982}]{av82}
{Azzopardi} M.,  {Vigneau} J.,  1982, \aas, 50, 291

\bibitem[\protect\citeauthoryear{{Baade}, {Rivinius}, {{\v S}tefl} \&
  {Kaufer}}{{Baade} et~al.}{2002}]{baa02}
{Baade} D.,  {Rivinius} T.,  {{\v S}tefl} S.,    {Kaufer} A.,  2002, \aap, 383,
  L31

\bibitem[\protect\citeauthoryear{{Balona} \& {Crampton}}{{Balona} \&
  {Crampton}}{1974}]{bc74}
{Balona} L.,  {Crampton} D.,  1974, \mnras, 166, 203

\bibitem[\protect\citeauthoryear{{Bica} \& {Dutra}}{{Bica} \&
  {Dutra}}{2000}]{bd00}
{Bica} E.,  {Dutra} C.~M.,  2000, \aj, 119, 1214

\bibitem[\protect\citeauthoryear{{Bica} \& {Schmitt}}{{Bica} \&
  {Schmitt}}{1995}]{bs95}
{Bica} E. L.~D.,  {Schmitt} H.~R.,  1995, \apjs, 101, 41

\bibitem[\protect\citeauthoryear{{Blair} \& {Gilmore}}{{Blair} \&
  {Gilmore}}{1982}]{bg82}
{Blair} M.,  {Gilmore} G.,  1982, \pasp, 94, 742

\bibitem[\protect\citeauthoryear{{Chromey} \& {Hasselbacher}}{{Chromey} \&
  {Hasselbacher}}{1996}]{ch96}
{Chromey} F.~R.,  {Hasselbacher} D.~A.,  1996, \pasp, 108, 944

\bibitem[\protect\citeauthoryear{{Cornett}, {Greason}, {Hill}, {Parker} \&
  {Waller}}{{Cornett} et~al.}{1997}]{uitcat97}
{Cornett} R.~H.,  {Greason} M.~R.,  {Hill} J.~K.,  {Parker} J.~W.,    {Waller}
  W.~H.,  1997, \aj, 113, 1011

\bibitem[\protect\citeauthoryear{{Cowley}, {Cowley}, {Jaschek} \&
  {Jaschek}}{{Cowley} et~al.}{1969}]{ccjj}
{Cowley} A.,  {Cowley} C.,  {Jaschek} M.,    {Jaschek} C.,  1969, \aj, 74, 375

\bibitem[\protect\citeauthoryear{{Evans} \& {Howarth}}{{Evans} \&
  {Howarth}}{2003}]{eh03}
{Evans} C.~J.,  {Howarth} I.~D.,  2003, \mnras, 345, 1223

\bibitem[\protect\citeauthoryear{{Fehrenbach} \& {Duflot}}{{Fehrenbach} \&
  {Duflot}}{1972}]{fd72}
{Fehrenbach} C.,  {Duflot} M.,  1972, \aap, 21, 321

\bibitem[\protect\citeauthoryear{{Fitzgerald}}{{Fitzgerald}}{1970}]{fg70}
{Fitzgerald} M.~P.,  1970, \aap, p.~234

\bibitem[\protect\citeauthoryear{{Garmany}, {Conti} \& {Massey}}{{Garmany}
  et~al.}{1987}]{gar87}
{Garmany} C.~D.,  {Conti} P.~S.,    {Massey} P.,  1987, \aj, 93, 1070

\bibitem[\protect\citeauthoryear{{Gilmore} \& {Howell}}{{Gilmore} \&
  {Howell}}{1998}]{imfconf}
{Gilmore} G.,  {Howell} D.,  eds, 1998, The Stellar Initial Mass Function, 38th
  Herstmonceux Conference.
A.S.P. Conference Series, Vol. 142

\bibitem[\protect\citeauthoryear{{Gray} \& {Garrison}}{{Gray} \&
  {Garrison}}{1987}]{gg87}
{Gray} R.~O.,  {Garrison} R.~F.,  1987, \apjs, 65, 581

\bibitem[\protect\citeauthoryear{{Gray} \& {Garrison}}{{Gray} \&
  {Garrison}}{1989}]{gg89b}
{Gray} R.~O.,  {Garrison} R.~F.,  1989, \apjs, 70, 623

\bibitem[\protect\citeauthoryear{{Gray}, {Napier} \& {Winkler}}{{Gray}
  et~al.}{2001}]{gnw01}
{Gray} R.~O.,  {Napier} M.~G.,    {Winkler} L.~I.,  2001, \aj, 121, 2148

\bibitem[\protect\citeauthoryear{{Harries}, {Hilditch} \& {Howarth}}{{Harries}
  et~al.}{2003}]{hhh03}
{Harries} T.~J.,  {Hilditch} R.~W.,    {Howarth} I.~D.,  2003, \mnras, 339, 157

\bibitem[\protect\citeauthoryear{{Henize}}{{Henize}}{1956}]{hen56}
{Henize} K.~G.,  1956, \apjs, 2, 315

\bibitem[\protect\citeauthoryear{{Heydari-Malayeri}}{{Heydari-Malayeri}}{1990}%
]{hey90}
{Heydari-Malayeri} M.,  1990, \aap, 234, 233

\bibitem[\protect\citeauthoryear{{Hodge}}{{Hodge}}{1985}]{h85}
{Hodge} P.~W.,  1985, \pasp, 97, 530

\bibitem[\protect\citeauthoryear{{Hummel}, {Szeifert}, {G{\" a}ssler},
  {Muschielok}, {Seifert}, {Appenzeller} \& {Rupprecht}}{{Hummel}
  et~al.}{1999}]{humm99}
{Hummel} W.,  {Szeifert} T.,  {G{\" a}ssler} W.,  {Muschielok} B.,  {Seifert}
  W.,  {Appenzeller} I.,    {Rupprecht} G.,  1999, \aap, 352, L31

\bibitem[\protect\citeauthoryear{{Humphreys}}{{Humphreys}}{1983}]{h83}
{Humphreys} R.~M.,  1983, \apj, 265, 176

\bibitem[\protect\citeauthoryear{{Humphreys}, {Kudritzki} \&
  {Groth}}{{Humphreys} et~al.}{1991}]{hkg91}
{Humphreys} R.~M.,  {Kudritzki} R.~P.,    {Groth} H.~G.,  1991, \aap, 245, 593

\bibitem[\protect\citeauthoryear{{Humphreys} \& {McElroy}}{{Humphreys} \&
  {McElroy}}{1984}]{hm84}
{Humphreys} R.~M.,  {McElroy} D.~B.,  1984, \apj, 284, 565

\bibitem[\protect\citeauthoryear{{Hutchings}}{{Hutchings}}{1966}]{hut66}
{Hutchings} J.~B.,  1966, \mnras, 132, 433

\bibitem[\protect\citeauthoryear{{Jaschek} \& {Jaschek}}{{Jaschek} \&
  {Jaschek}}{1990}]{jj}
{Jaschek} C.,  {Jaschek} M.,  1990, The Classification of Stars.
Cambridge University Press

\bibitem[\protect\citeauthoryear{{Keller}, {Wood} \& {Bessell}}{{Keller}
  et~al.}{1999}]{kwb99}
{Keller} S.~C.,  {Wood} P.~R.,    {Bessell} M.~S.,  1999, \aaps, 134, 489

\bibitem[\protect\citeauthoryear{{Landolt}}{{Landolt}}{1992}]{lan92}
{Landolt} A.~U.,  1992, \aj, 104, 372

\bibitem[\protect\citeauthoryear{{Lennon}}{{Lennon}}{1997}]{djl97}
{Lennon} D.~J.,  1997, \aap, 317, 871

\bibitem[\protect\citeauthoryear{{Lewis}, {Cannon}, {Taylor}, {Glazebrook},
  {Waller}, {Whittard}, {Wilcox} \& {Willis}}{{Lewis} et~al.}{2002}]{lewis02}
{Lewis} I.~J.,  {Cannon} R.~D.,  {Taylor} K.,  {Glazebrook} K.,  {Waller}
  L.~G.,  {Whittard} J.~D.,  {Wilcox} J.~K.,    {Willis} K.~C.,  2002, \mnras,
  333, 279

\bibitem[\protect\citeauthoryear{{Lindsay}}{{Lindsay}}{1961}]{l61}
{Lindsay} E.~M.,  1961, \aj, 66, 169

\bibitem[\protect\citeauthoryear{{Massey}}{{Massey}}{2002}]{pm02}
{Massey} P.,  2002, \apjs, 141, 81

\bibitem[\protect\citeauthoryear{{Massey} \& {Duffy}}{{Massey} \&
  {Duffy}}{2001}]{pm01}
{Massey} P.,  {Duffy} A.~S.,  2001, \apj, 550, 713

\bibitem[\protect\citeauthoryear{{Massey}, {Lang}, {DeGioia-Eastwood} \&
  {Garmany}}{{Massey} et~al.}{1995}]{pm95}
{Massey} P.,  {Lang} C.~C.,  {DeGioia-Eastwood} K.,    {Garmany} C.,  1995,
  \apj, 438, 188

\bibitem[\protect\citeauthoryear{{Meyssonnier} \& {Azzopardi}}{{Meyssonnier} \&
  {Azzopardi}}{1993}]{ma93}
{Meyssonnier} N.,  {Azzopardi} M.,  1993, \aaps, 102, 451

\bibitem[\protect\citeauthoryear{{Morgan} \& {Keenan}}{{Morgan} \&
  {Keenan}}{1973}]{mk73}
{Morgan} D.~H.,  {Keenan} P.~C.,  1973, \araap, 11, 29

\bibitem[\protect\citeauthoryear{{Morgan}, {Vassiliadis} \& {Dopita}}{{Morgan}
  et~al.}{1991}]{mvd91}
{Morgan} D.~H.,  {Vassiliadis} E.,    {Dopita} M.~A.,  1991, \mnras, 251, 51

\bibitem[\protect\citeauthoryear{{Morgan}, {Code} \& {Whitford}}{{Morgan}
  et~al.}{1955}]{mcw55}
{Morgan} W.~W.,  {Code} A.~D.,    {Whitford} A.~E.,  1955, \apjs, 2, 41

\bibitem[\protect\citeauthoryear{{Morgan}, {Harris} \& {Johnson}}{{Morgan}
  et~al.}{1953}]{mhj53}
{Morgan} W.~W.,  {Harris} D.~L.,    {Johnson} H.~L.,  1953, \apj, 118, 92

\bibitem[\protect\citeauthoryear{{Morgan}, {Keenan} \& {Kellman}}{{Morgan}
  et~al.}{1943}]{mkk}
{Morgan} W.~W.,  {Keenan} P.~C.,    {Kellman} E.,  1943, An atlas of stellar
  spectra.
Chicago Univ. Press

\bibitem[\protect\citeauthoryear{{Morgan} \& {Roman}}{{Morgan} \&
  {Roman}}{1950}]{mr50}
{Morgan} W.~W.,  {Roman} N.~G.,  1950, \apj, 112, 362

\bibitem[\protect\citeauthoryear{{Naz$\acute{\rm e}$}, {Vreux} \&
  {Rauw}}{{Naz$\acute{\rm e}$} et~al.}{2001}]{yael}
{Naz$\acute{\rm e}$} Y.,  {Vreux} J.-M.,    {Rauw} G.,  2001, A\&A, 372, 195

\bibitem[\protect\citeauthoryear{{Osmer} \& {Hiltner}}{{Osmer} \&
  {Hiltner}}{1974}]{oh74}
{Osmer} P.,  {Hiltner} W.~A.,  1974, \apj, 188, L5

\bibitem[\protect\citeauthoryear{{Parker}}{{Parker}}{2002}]{parker02}
{Parker} J.~W.,  2002, in {Crowther} P.~A.,  ed., The Earliest Stages of
  Massive Star Birth. A.S.P. Conference Series, Vol. 267, p.~401

\bibitem[\protect\citeauthoryear{{Parker}, {Hill}, {Cornett}, {Hollis},
  {Zamkoff}, {Bohlin}, {O'Connell}, {Neff}, {Roberts}, {Smith} \&
  {Stecher}}{{Parker} et~al.}{1998}]{parker98}
{Parker} J.~W.,  {Hill} J.~K.,  {Cornett} R.~H.,  {Hollis} J.,  {Zamkoff} E.,
  {Bohlin} R.~C.,  {O'Connell} R.~W.,  {Neff} S.~G.,  {Roberts} M.~S.,  {Smith}
  A.~M.,    {Stecher} T.~P.,  1998, \aj, 116, 180

\bibitem[\protect\citeauthoryear{{Pietrzynski}, {Udalski}, {Kubiak},
  {Szymanski}, {Wozniak} \& {Zebrun}}{{Pietrzynski} et~al.}{1998}]{puk98}
{Pietrzynski} G.,  {Udalski} A.,  {Kubiak} M.,  {Szymanski} M.,  {Wozniak} P.,
    {Zebrun} K.,  1998, Acta Astronomica, 48, 175

\bibitem[\protect\citeauthoryear{{Salpeter}}{{Salpeter}}{1955}]{s55}
{Salpeter} E.~E.,  1955, \apj, 121, 161

\bibitem[\protect\citeauthoryear{{Sanduleak}}{{Sanduleak}}{1968}]{sa68}
{Sanduleak} N.,  1968, \apj, 73, 246

\bibitem[\protect\citeauthoryear{{Sanduleak}}{{Sanduleak}}{1972}]{sa72}
{Sanduleak} N.,  1972, \aap, 17, 326

\bibitem[\protect\citeauthoryear{{Schmidt-Kaler}}{{Schmidt-Kaler}}{1982}]{sk82}
{Schmidt-Kaler} T.,  1982, in {Schaifers} K.,  {Voigt} H.~H.,  eds,
  Landolt-B\"{o}rnstein, Group VI, Vol 2b Springer-Verlag, p.~1

\bibitem[\protect\citeauthoryear{{Slettebak}}{{Slettebak}}{1954}]{slet54}
{Slettebak} A.,  1954, 1954, 119, 146

\bibitem[\protect\citeauthoryear{{Udalski}, {Szyma${\rm \acute{n}}$ski},
  {Kubiak}, {Pietrzy${\rm \acute{n}}$ski}, {Wo${\rm \acute{z}}$niak} \& {${\rm
  \dot{Z}}$ebru${\rm \acute{n}}$}}{{Udalski} et~al.}{1998}]{u98}
{Udalski} A.,  {Szyma${\rm \acute{n}}$ski} M.,  {Kubiak} M.,  {Pietrzy${\rm
  \acute{n}}$ski} G.,  {Wo${\rm \acute{z}}$niak} P.,    {${\rm
  \dot{Z}}$ebru${\rm \acute{n}}$} K.,  1998, Acta Astron., 48, 147

\bibitem[\protect\citeauthoryear{{van der Hucht}}{{van der
  Hucht}}{1996}]{vdh96}
{van der Hucht} K.~A.,  1996, in {Vreux} J.~M.,  ed., Wolf-Rayet Stars in the
  Framework of Stellar Evolution; $33^{\rm rd}$ Li$\grave{e}$ge Int. Astroph.
  Coll. p.~1

\bibitem[\protect\citeauthoryear{{Walborn}}{{Walborn}}{1972}]{wal72}
{Walborn} N.~R.,  1972, AJ, 77, 312

\bibitem[\protect\citeauthoryear{{Walborn}}{{Walborn}}{1973}]{wal73}
{Walborn} N.~R.,  1973, \aj, 78, 1067

\bibitem[\protect\citeauthoryear{{Walborn}}{{Walborn}}{1977}]{wal77}
{Walborn} N.~R.,  1977, \apj, 215, 53

\bibitem[\protect\citeauthoryear{{Walborn} \& {Fitzpatrick}}{{Walborn} \&
  {Fitzpatrick}}{1990}]{wf90}
{Walborn} N.~R.,  {Fitzpatrick} E.~L.,  1990, \pasp, 102, 379

\bibitem[\protect\citeauthoryear{{Walborn}, {Howarth}, {Herrero} \&
  {Lennon}}{{Walborn} et~al.}{2003}]{wh03}
{Walborn} N.~R.,  {Howarth} I.~D.,  {Herrero} A.,    {Lennon} D.~J.,  2003,
  ApJ, 588, 1025

\bibitem[\protect\citeauthoryear{{Walborn}, {Lennon}, {Heap}, {Lindler},
  {Smith}, {Evans} \& {Parker}}{{Walborn} et~al.}{2000}]{wal00}
{Walborn} N.~R.,  {Lennon} D.~J.,  {Heap} S.,  {Lindler} D.~J.,  {Smith} L.~J.,
   {Evans} C.~J.,    {Parker} J.~W.,  2000, \pasp, 112, 1243

\bibitem[\protect\citeauthoryear{{Wegner}}{{Wegner}}{2000}]{weg00}
{Wegner} W.,  2000, \mnras, 319, 771

\bibitem[\protect\citeauthoryear{{Zaritsky}, {Harris}, {Thompson}, {Grebel} \&
  {Massey}}{{Zaritsky} et~al.}{2002}]{z02}
{Zaritsky} D.,  {Harris} J.,  {Thompson} I.~B.,  {Grebel} E.~K.,    {Massey}
  P.,  2002, \aj, 123, 855

\end{thebibliography}

\appendix
\section{Catalogue description}
\label{app_cat}

The full catalogue of spectral types for the 2dF sample is available
electronically through the on-line edition of Monthly Notices, and at
the Centre de Donn\'{e}es astronomiques de Strasbourg
(CDS\footnote{\tt{ftp://cdsarc.u-strasbg.fr/pub/cats/J/MNRAS/XXX/YYY}}).
The data are assembled in two tables, one containing the basic
observational data from the spectroscopic survey, and one containing
cross-identifications.

Table~\ref{catalogue} shows an extract from the data catalogue, to
illustrate the format.  For convenience, we assign a 4-digit 2dFS
number to each star; entries 1--4054 are the main, APM-selected,
targets, sorted by RA, and entries 5001--5107 are the UIT-selected
targets, separately sorted by RA.  There are no catalogue entries
4055--5000.  For each entry we give the adopted co-ordinates and $B$
magnitude (Section~\ref{photom_txt}).  Co-ordinates are quoted from
the sources used for spectrograph fibre configuration, i.e., APM for
2dFS entries 1--4054, OGLE for entries 5001--5107
(cf. \citealt{hhh03}).  We then list the adopted 2dF spectral type,
and notes on Balmer-line emission for \Ha, \Hb, and \Hg, as available
(Section~\ref{Bem_stat}).  The overall distribution of spectral types
is summarized in Table~\ref{sp_summary}.  Note that the luminosity
classes for the A0 and B-type stars are given in parentheses, to
reflect the fact that they are not based solely on morphological
considerations and therefore are not strict `MK' types.

The supplementary catalogue of cross-identifications includes full
photometric results from Massey, OGLE, and MCPS
(Section~\ref{sec_phot}), and cross-identifications with the UIT,
Sanduleak, AzV, and MA93 catalogues (\citealt{uitcat97};
\citealt{sa68, sa72}; \citealt{av82};
\citealt{ma93}).
The catalogue also includes pertinent information from the catalogue
of SMC clusters by \citet{bd00}, which is, essentially, the catalogue
of
\citet{bs95}, with the inclusion of new clusters identified in the
OGLE survey \citep{puk98}.  Our purpose in including this information
is to provide an indication of whether a given star is likely to be a
cluster or association member, or a field star (relevant to the
interpretation of the HRD).

There are 49 Sk stars matched in the 2dFS catalogue, 172 AzV stars,
219 MA93 stars, and 454 UIT stars (including 107 UIT-selected targets).
These correspondences are to be understood to be simply positional
matches for the photometric, UIT, and MA93 catalogues (which list
co-ordinates to the nearest arcsecond or better), but are actual
cross-identifications for the Sk and AzV catalogues, using published
finder charts.

\begin{table*}
\begin{center}
\caption[] {An illustrative section of the on-line 2dF catalogue.
Sources of adopted $B$ magnitudes are APM, Massey, OGLE, and Zaritsky
(coded A, M, O, Z; see Section~\ref{photom_txt}).  The `Balmer notes'
summarize lines in blue (B: g=\Hg, b=\Hb) and red (R: \Ha) spectra,
coded as a (absorption); n (narrow [nebular?] emission); r (resolved
[but narrow] emission); d (double-peaked emission); p (P-Cyg profile).}
\label{catalogue}
\begin{tabular}{ccccccccclrl}
\hline
2dFS\# & \multicolumn{3}{c}{$\alpha$ (J2000)} & 
\multicolumn{3}{c|}{$\delta$ (J2000)} 
&$B$
&Source
&\multicolumn{1}{c}{Spectral type} 
&\multicolumn{2}{c}{Balmer notes} \\
\hline
 0600 & 00 & 46 & 23.13 & $-$72 & 50 & 17.4 &  16.36 &  O &   A3$\;$II        &            B[ga] &   R[n] \\
 0601 & 00 & 46 & 28.15 & $-$72 & 52 & 23.3 &  16.63 &  O &   B8$\;$(III)     &         B[ga bn] &         \\
 0602 & 00 & 46 & 28.40 & $-$73 & 27 & 43.0 &  15.15 &  Z &   B1--5$\;$(II)   &            B[gn] &   R[n] \\
 0603 & 00 & 46 & 30.53 & $-$72 & 26 & 58.4 &  16.84 &  Z &   B1--5$\;$(IV)   &            B[ga] &         \\
 0604 & 00 & 46 & 32.90 & $-$72 & 25 & 16.8 &  16.53 &  Z &   B0.5$\;$(V)     &            B[ga] &         \\
 0605 & 00 & 46 & 33.08 & $-$72 & 10 & 39.3 &  16.86 &  Z &   B0--5$\;$(V)    &            B[ga] &         \\
 0606 & 00 & 46 & 33.47 & $-$73 & 39 & 25.8 &  13.34 &  M &   B8$\;$(Ib)      &         B[ga ba] &         \\
 0607 & 00 & 46 & 33.86 & $-$73 & 53 & 04.7 &  17.60 &  Z &   B0--5$\;$(V)    &         B[ga ba] &         \\
 0608 & 00 & 46 & 35.12 & $-$72 & 56 & 04.2 &  15.93 &  O &   B3$\;$(III)     &         B[ga ba] &         \\
 0609 & 00 & 46 & 38.10 & $-$73 & 55 & 13.7 &  14.31 &  Z &   B0.5$\;$(IV)    &            B[ga] &         \\
 0610 & 00 & 46 & 40.19 & $-$73 & 31 & 16.9 &  14.17 &  Z &   O7--8$\;$V      &         B[gn bn] &         \\
 0611 & 00 & 46 & 41.70 & $-$74 & 00 & 49.0 &  17.11 &  Z &   B0--5$\;$(V)    &         B[gn bn] &         \\
 0612 & 00 & 46 & 43.21 & $-$73 & 28 & 03.4 &  16.46 &  O &   B1--5$\;$(III)  &         B[ga bn] &   R[n] \\
 0613 & 00 & 46 & 46.20 & $-$73 & 41 & 13.2 &  16.35 &  Z &   B1--5$\;$(III)  &            B[ga] &         \\
 0614 & 00 & 46 & 47.00 & $-$73 & 58 & 29.7 &  17.42 &  Z &   A3$\;$II        &         B[ga ba] &         \\
 0615 & 00 & 46 & 47.38 & $-$72 & 33 & 46.7 &  17.37 &  Z &   F8              &            B[g ] &         \\
 0616 & 00 & 46 & 47.72 & $-$72 & 58 & 43.8 &  17.10 &  Z &   B0--5$\;$(V)    &            B[ga] &   R[n] \\
 0617 & 00 & 46 & 48.07 & $-$73 & 57 & 10.6 &  16.88 &  Z &   A2$\;$III       &         B[ga ba] &         \\
 0618 & 00 & 46 & 48.95 & $-$73 & 30 & 37.3 &  16.66 &  O &   B8--A0$\;$(II)  &         B[gn bn] &         \\
 0619 & 00 & 46 & 50.81 & $-$72 & 17 & 10.9 &  15.78 &  Z &   B0--5$\;$(II)   &            B[gr] &         \\
 0620 & 00 & 46 & 50.88 & $-$73 & 55 & 21.5 &  14.57 &  Z &   B2$\;$(III)     &            B[ga] &         \\
\hline
\end{tabular}                                                                         
\end{center}
\end{table*}

\end{document}